\documentclass[10pt,twocolumn,amssymb,amsmath,aps,pra]{revtex4-1}
\usepackage{graphicx}

\begin{document}

\title{Mass-polariton theory of light in dispersive media}
\date{December 26, 2017}
\author{Mikko Partanen}
\affiliation{Engineered Nanosystems Group, School of Science, Aalto University, P.O. Box 11000, 00076 Aalto, Finland}
\author{Jukka Tulkki}
\affiliation{Engineered Nanosystems Group, School of Science, Aalto University, P.O. Box 11000, 00076 Aalto, Finland}

\begin{abstract}
We have recently shown that the electromagnetic pulse in a medium is made of
mass-polariton (MP) quasiparticles, which are quantized coupled states of the field and
an atomic mass density wave (MDW)
[M.~Partanen \emph{et al.}, Phys.~Rev.~A \textbf{95}, 063850 (2017)]. 
In this work, we generalize the MP theory of light for dispersive media
assuming that absorption and scattering losses are very small.
Following our previous work, we present two different approaches
to the coupled state of light: (1) the MP quasiparticle theory, which
is derived by only using the fundamental conservation laws and the
Lorentz transformation; (2) the classical optoelastic continuum dynamics (OCD),
which is a generalization of the electrodynamics of continuous media
to include the dynamics of the medium under the influence of optical forces. 
We show that the total momentum and the transferred mass of the
light pulse can be determined in a straightforward way if we know the
field energy of the pulse and the dispersion relation of the medium.
In analogy to the nondispersive case, we also find unambiguous
correspondence between the MP and OCD theories.
For the coupled MP state of a single photon and the medium,
we obtain the total MP momentum $p_\mathrm{MP}=n_\mathrm{p}\hbar\omega/c$,
where $n_\mathrm{p}$ is the phase refractive index.
The field's share of the MP momentum is equal to $p_\mathrm{field}=\hbar\omega/(n_\mathrm{g}c)$,
where $n_\mathrm{g}$ is the group refractive index
and the share of the MDW is equal to $p_\mathrm{MDW}=p_\mathrm{MP}-p_\mathrm{field}$. Thus, as in a nondispersive medium,
the total momentum of the MP is equal to the
Minkowski momentum and the field's share
of the momentum is equal to the Abraham momentum.
We also show that the correspondence between the MP and OCD models
and the conservation of momentum at
interfaces  gives an unambiguous formula for the optical force.
The dynamics of the light pulse and the related MDW
lead to nonequilibrium of the medium and to relaxation
of the atomic density by sound waves in the same way as
for nondispersive media.
We also carry out simulations for optimal measurements of atomic displacements
related to the MDW in silicon. In the simulations, we consider different waveguide cross sections
and optical pulse widths and account for the breakdown threshold irradiance of materials.
We also compare the MP theory to previous theories of the momentum of light
in a dispersive medium.
We show that our generalized MP theory resolves all the
problems related to the Abraham-Minkowski dilemma in a dispersive
medium.
\end{abstract}

\maketitle

\section{Introduction}

\vspace{-0.2cm}

Previous theories of light in a medium have neglected the possibility of
an associated mass density wave (MDW) formed by small atomic movements
caused by the optical force that is alternately accelerating and decelerating medium atoms.
We have recently shown that the MDW is an unavoidable part
of the consistent theory of light in a medium \cite{Partanen2017c}.
In the single-photon picture, the coupling of the electromagnetic
field to the atomic MDW gives rise to mass-polariton (MP) quasiparticles,
which are covariant coupled states of the field and matter having
a nonzero rest mass \cite{Partanen2017c}.
The coupled state of the field and matter can also be described
by using classical optoelastic continuum dynamics (OCD) \cite{Partanen2017c}.
In the OCD model, the electrodynamics of
continuous media \cite{Landau1984} is generalized to include the coupling
between the field and matter and the related continuum dynamics of the medium.

Accounting for the MDW coupled to the electromagnetic field, the photon mass drag effect
has been shown \cite{Partanen2017c} to resolve the centennial
Abraham-Minkowski controversy of optical momentum in a medium
\cite{Leonhardt2006,Cho2010,Pfeifer2007,Barnett2010b,Barnett2010a,Leonhardt2014,Saldanha2017,Brevik2017,Ramos2015,Crenshaw2013,Mansuripur2010,Kemp2011}.
This controversy has its origin in the formulation of two rivaling
momentum densities for light by Abraham,
$\mathbf{g}_\mathrm{A}=\mathbf{E}\times\mathbf{H}/c^2$
\cite{Abraham1909,Abraham1910}, and by Minkowski,
$\mathbf{g}_\mathrm{M}=\mathbf{D}\times\mathbf{B}$ \cite{Minkowski1908},
where $c$ is the speed of light in vacuum, $\mathbf{E}$ and $\mathbf{H}$
are the electric- and magnetic-field strengths, and $\mathbf{D}$ and
$\mathbf{B}$ are the electric and magnetic
flux densities. For a nondispersive medium, the momentum
densities $\mathbf{g}_\mathrm{A}$ and
$\mathbf{g}_\mathrm{M}$ correspond to the single-photon momenta
$p_\mathrm{A}=\hbar\omega/(nc)$ or $p_\mathrm{M}=n\hbar\omega/c$,
respectively, where $\hbar$ is the reduced Planck constant, $\omega$ is the angular
frequency of the field, and $n$ is the refractive index of the medium.
In order to determine the momentum of light in a medium,
several experimental setups have been introduced
\cite{Campbell2005,Sapiro2009,Jones1954,Jones1978,Walker1975,She2008,Zhang2015,Astrath2014,Ashkin1973,Casner2001,Brevik1979}
but with partly controversial results.
In the recently developed MP theory \cite{Partanen2017c}, the Abraham momentum $p_\mathrm{A}$ is
the momentum of the electromagnetic field of the coupled MP state
while the difference $p_\mathrm{M}-p_\mathrm{A}$ is carried by the MDW.
The total MP momentum is then of the Minkowski
form $p_\mathrm{M}=n\hbar\omega/c$.

The initial derivation of the MP theory in Ref.~\cite{Partanen2017c}
assumed a nondispersive medium. In this work, we generalize
the MP quasiparticle model based on the conservation laws and
the Lorentz transformation for dispersive media.
Following Ref.~\cite{Partanen2017c}, we also present
the complementary classical OCD model, which we have
generalized for a dispersive medium.
The OCD model uses the optoelastic force density to calculate
the coupled Newtonian dynamics of the field and the medium
\cite{Partanen2017c}. The calculations show that the quasiparticle
and continuum dynamics models are in full agreement in the limit of a monochromatic
field, i.e., when the photon picture becomes reasonable, and also in the limit
of weak dispersion.

This paper is organized as follows: Section \ref{sec:experiments}
presents a brief summary of the most conclusive experiments to
measure the momentum of light in dispersive media. Section
\ref{sec:fieldsolution} reviews the well-known principles
of the dispersion relations and the solution of the electric
and magnetic fields of a light pulse. This is followed by
presenting the OCD model generalized for dispersive media
in Sec.~\ref{sec:ocdmodel} and the related complementary
MP quasiparticle model in Sec.~\ref{sec:quasiparticlemodel}.
In this work, we consider the OCD model first since it is for most readers
easier to approach being based on the familiar concepts of Maxwell's and
Newton's theories. However, the theories are independent and the reader
can also start from the MP quasiparticle model.
Section \ref{sec:simulations} presents the OCD
simulations of a Gaussian light pulse propagating in
linearly and nonlinearly dispersive media.
To facilitate the planning of possible experiments
of the transferred mass of the MDW, we also compute the atomic
displacements due to the MDW in a schematic silicon waveguide structure.
The results of the OCD and MP quasiparticle models
are compared in Sec.~\ref{sec:comparison}.
We also compare our theory to selected previous
experiments and theories that have been used to determine
the momentum of light in a dispersive medium.
Finally, conclusions are drawn in Sec.~\ref{sec:conclusions}.

\section{\label{sec:experiments}Brief summary of experiments}

\begin{figure}
\includegraphics[width=0.48\textwidth]{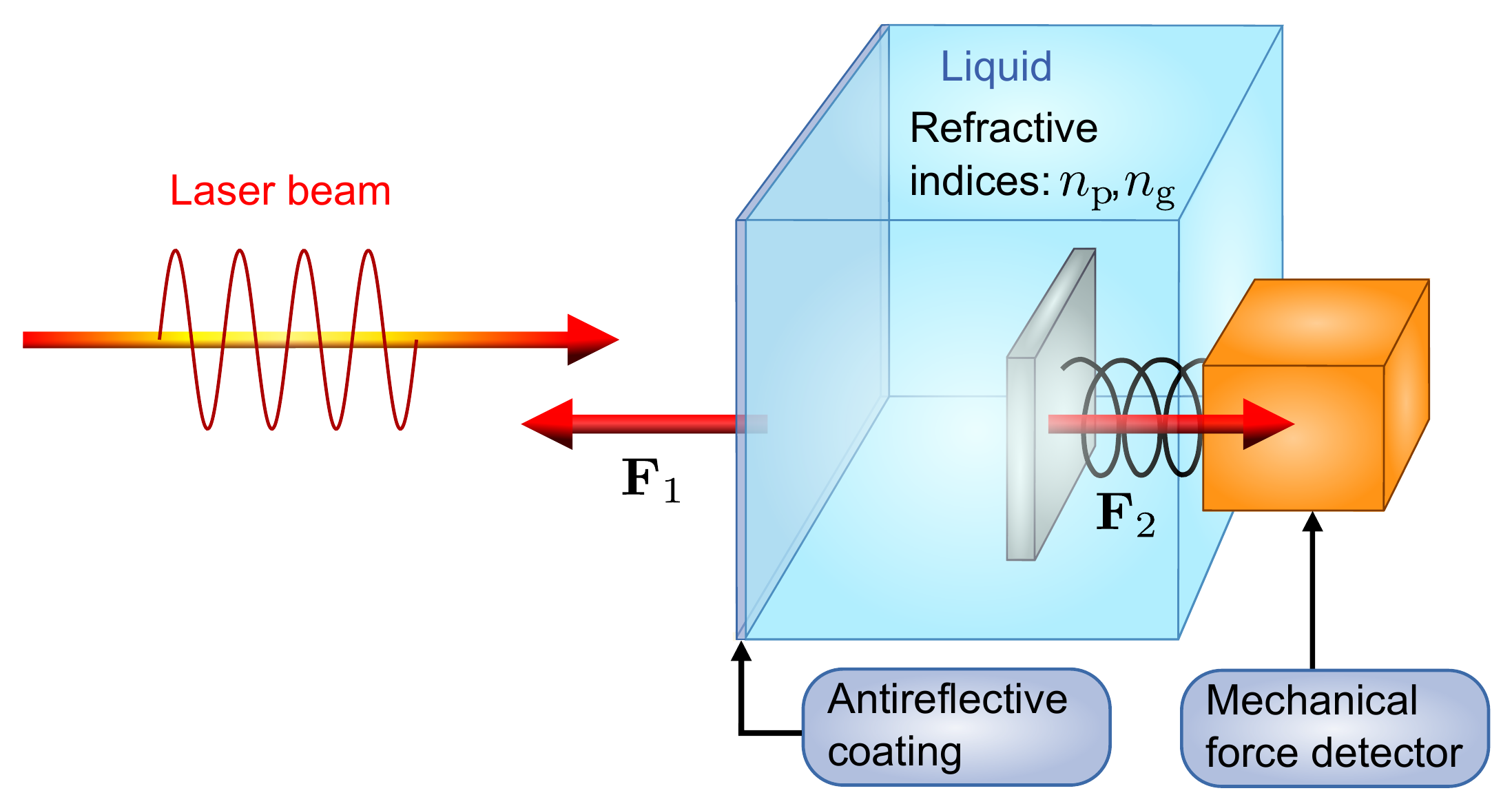}
\caption{\label{fig:experiment}
(Color online) Schematic illustration of an experimental setup for
the measurement of the electromagnetic forces due to a light beam
in a dispersive medium. Light enters from vacuum
to a liquid container with antireflective coating. Inside the liquid
having phase and group refractive indices $n_\mathrm{p}$ and $n_\mathrm{g}$,
light is fully reflected from a mirror attached to a detector that measures
the resulting force $\mathbf{F}_2$.
$\mathbf{F}_1$ is a recoil force that
balances the conservation law of momentum at the interface.}
\vspace{-0.1cm}
\end{figure}

The most conclusive set of experiments to measure
the momentum of light in a medium were started in 1954 by
Jones and Richards \cite{Jones1954} who studied the pressure exerted
by light on a reflector immersed in a liquid with known refractive index.
By performing the experiment with a number of liquids
of varying refractive index, they showed with 1.2\% precision that the pressure
on a reflector immersed in a liquid scales linearly with the refractive index.
The experiment was repeated in 1978 by Jones and Leslie \cite{Jones1978} with
0.05\% precision. A more than tenfold
improvement in precision was possible by using a laser
as a light source and multilayer reflectors of high reflectivity and low absorption.
The accuracy obtained
was sufficient to conclusively show that the
force on the mirror is directly proportional to the
phase refractive index $n_\mathrm{p}$ and not to the group
refractive index $n_\mathrm{g}$.
A principally identical schematic experimental setup is
illustrated in Fig.~\ref{fig:experiment}.
If the force $\mathbf{F}_2$ on the perfect lossless
reflector results from the single-photon
impulses $\Delta\mathbf{p}_i$ in time $\Delta t$,
we obtain $\mathbf{F}_2=\sum_i\Delta\mathbf{p}_i/\Delta t$.
Then the experiment unambiguously supports
the Minkowski formula $p=n_\mathrm{p}\hbar\omega/c$
\cite{Pfeifer2007,Barnett2010b,Barnett2010a},
provided that we know the intensity and the frequency
of a monochromatic laser beam.
There exist also other experiments that have been
interpreted to support either the Minkowski or Abraham
momentum \cite{Campbell2005,Sapiro2009,Walker1975,She2008,Zhang2015,Astrath2014,Ashkin1973,Casner2001}.
In these experiments, the relation of
the measured force or other quantity to the momentum of light
is much more subtle and analyzing these experiments using
our theory is a topic of a separate work.

If we consider light in a dispersive medium as a coupled state
of the field and matter, we are expected to be able to apply the
de Broglie wavelength in the analysis of diffraction experiments.
In the diffraction experiments, one obtains the de Broglie wavelength
which is related to the momentum of the coupled state as
$\lambda=h/p$,
where $h$ is the (nonreduced) Planck constant.
Since numerous diffraction experiments have confirmed that
the wavelength fulfilling the diffraction condition
is given by $\lambda=\lambda_0/n_\mathrm{p}$,
where $\lambda_0$ is the vacuum wavelength,
we obtain $p=n_\mathrm{p}\hbar\omega_0/c$,
which is again of the Minkowski form.
Note that so far there are no reported measurements
of the transferred mass of a light pulse.

\section{\label{sec:fieldsolution}Solution of fields in dispersive media}

\subsection{General dispersion}

In dispersive media, the phase velocity and the phase refractive index depend on frequency
$\omega(k)=ck/n_\mathrm{p}(\omega)$, where $n_\mathrm{p}(\omega)$ is
the frequency-dependent phase refractive index.
The phase velocity is given by $v_\mathrm{p}(\omega)=c/n_\mathrm{p}(\omega)=\omega(k)/k$
and the group velocity by the formula
$v_\mathrm{g}(\omega)=c/n_\mathrm{g}(\omega)=\partial\omega(k)/\partial k$,
where $n_\mathrm{g}(\omega)$ is the group refractive index.

The most general forms of the electric and magnetic fields
of a linearly polarized one-dimensional
light pulse propagating in $x$ direction in a dispersive
medium can be written as \cite{Griffiths1998}
\begin{equation}
 \mathbf{E}(\mathbf{r},t)
 =\mathrm{Re}\Big[\int_{-\infty}^\infty \tilde E(k)e^{i[kx-\omega(k) t]}dk\Big]\hat{\mathbf{y}},
 \label{eq:electricfieldgeneral}
\end{equation}
\vspace{-0.2cm}
\begin{equation}
 \mathbf{H}(\mathbf{r},t)
 =\mathrm{Re}\Big[\int_{-\infty}^\infty \tilde H(k)e^{i[kx-\omega(k) t]}dk\Big]\hat{\mathbf{z}},
 \label{eq:magneticfieldgeneral}
\end{equation}
where $\hat{\mathbf{y}}$ and $\hat{\mathbf{z}}$
are unit vectors with respect to $y$ and $z$ axes and
$\tilde E(k)$ and $\tilde H(k)$ are the Fourier components of the
electric and magnetic fields. The field components
are related to each other by
$\tilde H(k)=\sqrt{\varepsilon[\omega(k)]/\mu[\omega(k)]}\tilde E(k)$,
where $\varepsilon(\omega)$ and $\mu(\omega)$ are the frequency-dependent
permittivity and permeability of the medium.
These are related to the phase refractive index as
$\varepsilon(\omega)\mu(\omega)=\varepsilon_0\mu_0n_\mathrm{p}(\omega)^2$,
where $\varepsilon_0$ and $\mu_0$ are the permittivity and permeability
of the vacuum respectively. The electric and magnetic fields in Eqs.~\eqref{eq:electricfieldgeneral}
and \eqref{eq:magneticfieldgeneral} are exact solutions
of Maxwell's equations.

\vspace{-0.1cm}

\subsection{\label{sec:dispersionlinear}Linear dispersion}

We first investigate a light pulse in a dispersive medium, where the
dispersion relation is effectively linear near
the central frequency $\omega_0$
containing the first terms of the Taylor expansion of $\omega(k)$ as
\begin{equation}
\omega(k)\approx\omega_0+(c/n_\mathrm{g})(k-k_\mathrm{0,med}),
\label{eq:lineardispersion}
\end{equation}
where $k_\mathrm{0,med}=n_\mathrm{p}k_0$ is the wave number corresponding to $\omega_0$
in the medium, $k_0=\omega_0/c$ is the wave number in vacuum,
$n_\mathrm{p}=n_\mathrm{p}(\omega_0)$ is the phase refractive index for $\omega_0$,
and the group refractive index $n_\mathrm{g}$ is constant.
The linear dispersion relation in Eq.~\eqref{eq:lineardispersion} is a good approximation
for any general dispersion relation if
the frequency spread of the wave packet is relatively small,
the dispersion relation does not have sharp variations due to resonances,
and if the wave packet does not travel over very long distances. Otherwise,
higher-order terms in the Taylor expansion of $\omega(k)$ also become important.

For frequencies deviating from $\omega_0$,
the linear dispersion relation in Eq.~\eqref{eq:lineardispersion} defines the frequency-dependent phase refractive index as
\begin{equation}
 n_\mathrm{p}(\omega)=n_\mathrm{g}+(n_\mathrm{p}-n_\mathrm{g})\frac{\omega_0}{\omega}.
\end{equation}
The linear form of the dispersion relation in Eq.~\eqref{eq:lineardispersion} is
known to be the most general form of the dispersion relation, which
does not lead to the distortion of the pulse envelope while the pulse propagates.

We assume a Gaussian light pulse with
$\tilde E(k)=\tilde E_0e^{-[(k-n_\mathrm{p}k_0)/(n_\mathrm{p}\Delta k_0)]^2/2}$
where $\tilde E_0$ is a normalization factor
and $\Delta k_0$ is the standard deviation
of the wave number in vacuum related to the pulse width
in the $x$ direction as $\Delta x=1/(\sqrt{2}n_\mathrm{p}\Delta k_0)$.
The corresponding standard deviation in time is then
$\Delta t=n_\mathrm{p}\Delta x/c=1/(\sqrt{2}\Delta k_0c)$
and the full width at half maximum is $\Delta t_\mathrm{FWHM}=2\sqrt{2\ln 2}\,\Delta t$.
Using Eq.~\eqref{eq:electricfieldgeneral} and the linear dispersion
relation in Eq.~\eqref{eq:lineardispersion},
the electric field then becomes
\vspace{-0.1cm}
\begin{align}
 \mathbf{E}(\mathbf{r},t)
 & =\sqrt{2\pi}\,n_\mathrm{p}\Delta k_0\tilde E_0\cos\Big[n_\mathrm{p}k_0\Big(x-\frac{ct}{n_\mathrm{p}}\Big)\Big]\nonumber\\
 &\hspace{0.5cm}\times e^{-(n_\mathrm{p}\Delta k_0)^2(x-ct/n_\mathrm{g})^2/2}\hat{\mathbf{y}}.
 \label{eq:electricfield}
\end{align}
The normalization factor $\tilde E_0$ in Eq.~\eqref{eq:electricfield}
can be determined
so that the integral of the corresponding instantaneous
energy density over $x$ gives $U_0/A$,
where $A$ is a cross-sectional area and $U_0$ is the total
energy of the light pulse.

\begin{figure}
\centering
\includegraphics[width=\columnwidth]{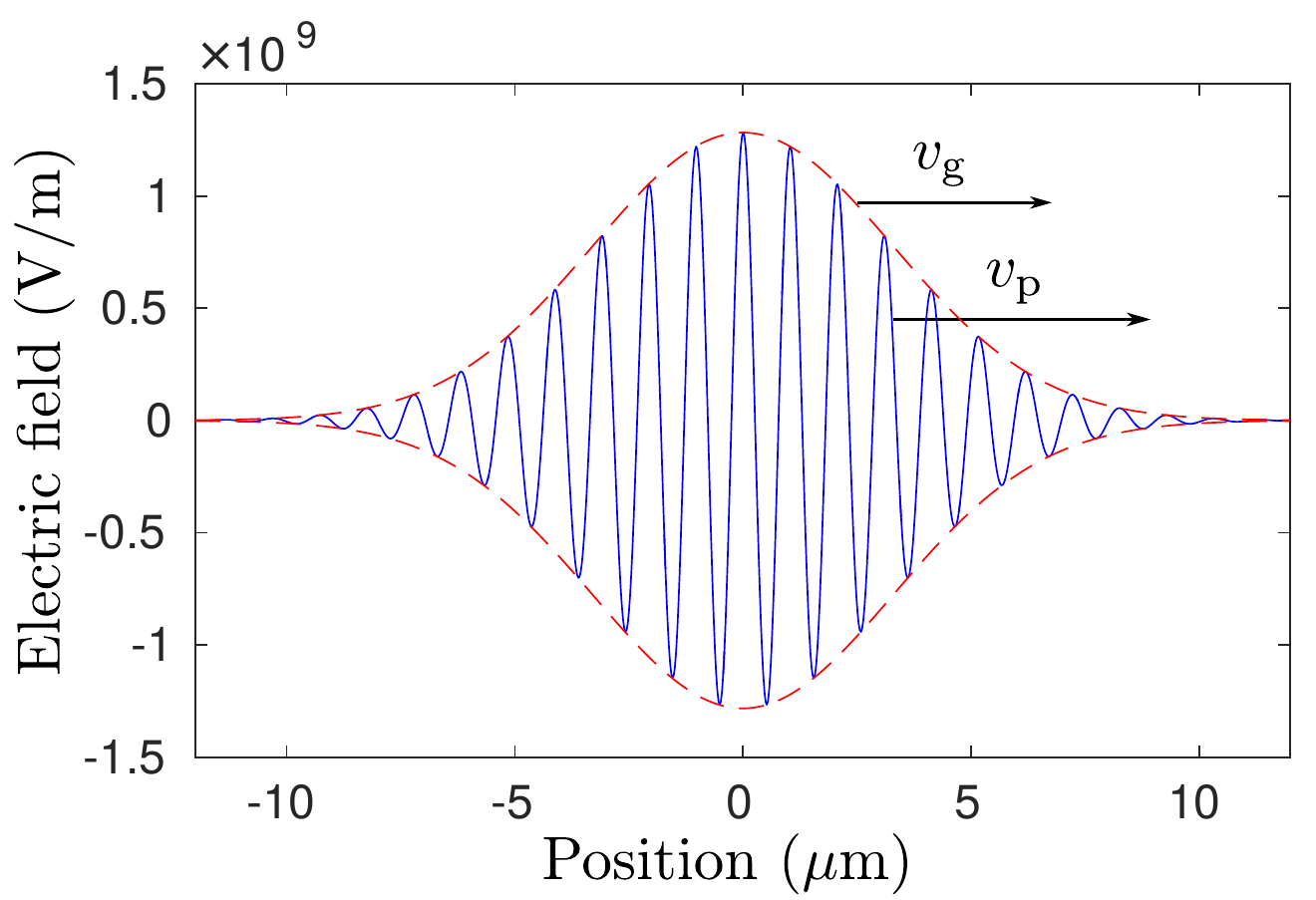}
\caption{\label{fig:efield}
(Color online) Example of the electric field and its envelope function in the case
of an ultrashort Gaussian light pulse
of vacuum wavelength $\lambda_0=1550$ nm, $\Delta t_\mathrm{FWHM}=27$ fs,
and energy $U_0=1$ $\mu$J per cross-sectional area of diameter $d=100$ $\mu$m.
The phase and group refractive indices for the central frequency in
a linearly dispersive medium are assumed to be $n_\mathrm{p}=1.5$ and $n_\mathrm{g}=2$.
The wave envelope propagates at the group velocity $v_\mathrm{g}=c/n_\mathrm{g}$
while the individual peaks and troughs inside the wave envelope propagate
at the phase velocity $v_\mathrm{p}=c/n_\mathrm{p}$.}
\vspace{-0.2cm}
\end{figure}

An example of the electric field of a Gaussian light pulse given in Eq.~\eqref{eq:electricfield}
is presented in Fig.~\ref{fig:efield}.
The envelope function described by the exponential factor in Eq.~\eqref{eq:electricfield}
propagates at the group velocity $v_\mathrm{g}=c/n_\mathrm{g}$ while the 
individual peaks and troughs inside the wave envelope propagate
at the phase velocity $v_\mathrm{p}=c/n_\mathrm{p}$.
In other words, the phase velocity $v_\mathrm{p}(\omega)$ describes the propagation
velocity of individual frequency components while the
amplitudes of the frequency components add up to produce
a wave packet, which propagates at the group velocity \cite{Griffiths1998}.
Therefore, the total energy of the wave packet propagates
at the group velocity.

\vspace{-0.2cm}

\subsection{\label{sec:dispersionnonlinear}Nonlinear dispersion}

\begin{figure}
\centering
\includegraphics[width=0.9\columnwidth]{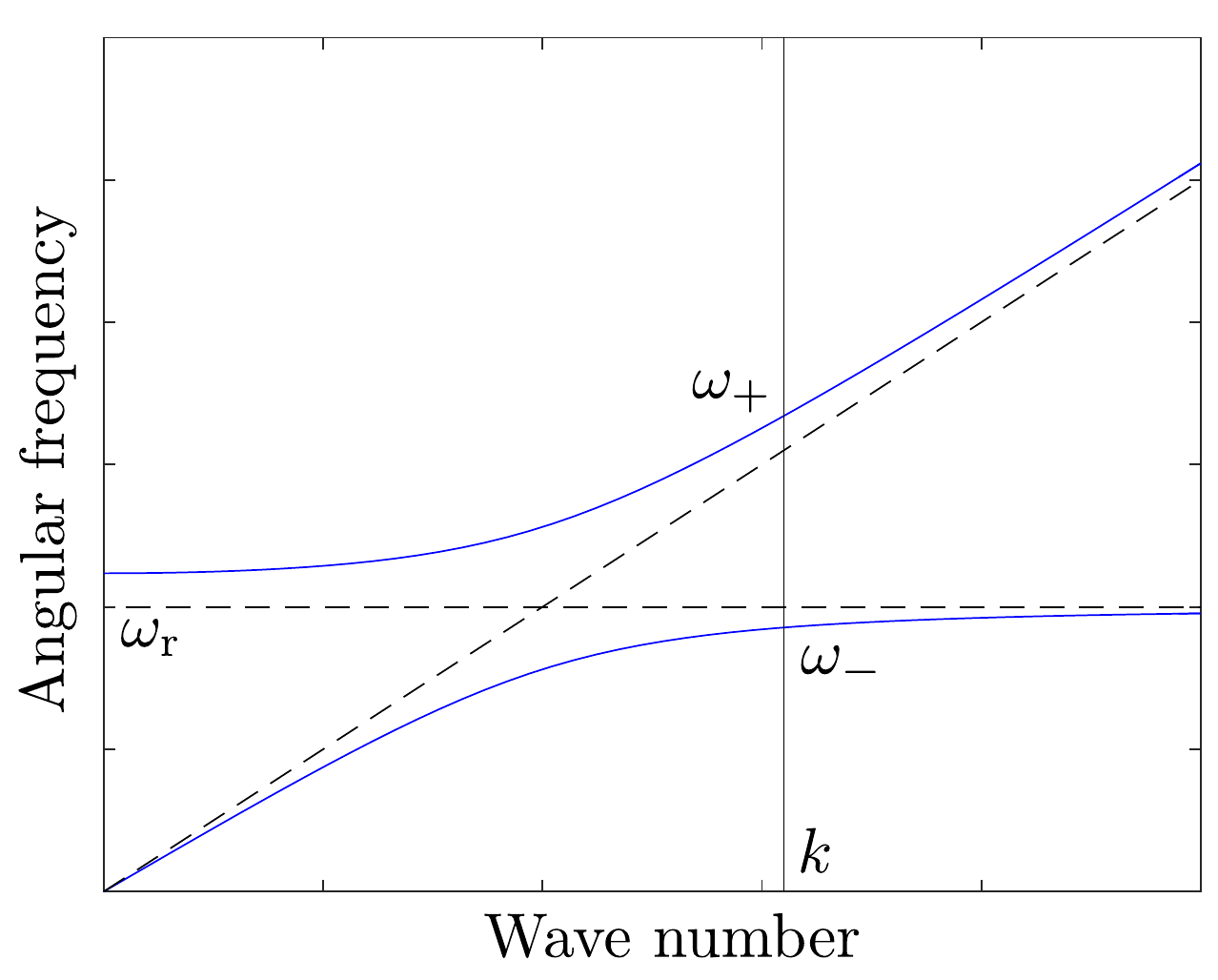}
\caption{\label{fig:branches}
(Color online) An example of a two-branch polariton dispersion curve with
a single resonance frequency $\omega_\mathrm{r}$.
For the lower branch $\omega<\omega_\mathrm{r}$ and for the upper branch $\omega>\omega_\mathrm{r}$.}
\end{figure}

In general, the linear dispersion relation above cannot provide
a complete description of dispersion close
to resonances or in the case of large frequency ranges. Therefore, 
following some previous works on the Abraham-Minkowski
controversy \cite{Milonni2010}, we study as an example of nonlinear dispersion
a simple Lorentz model for a dielectric medium with
a single resonance frequency $\omega_\mathrm{r}$ and zero damping factor \cite{Peiponen1999}.
The imaginary part of the refractive index can
be assumed zero at $\omega_0$ and the real part of the refractive index is given by \cite{Milonni2010}
\begin{equation}
 n_\mathrm{p}(\omega)=\sqrt{1+\frac{\omega_\mathrm{p}^2}{\omega_\mathrm{r}^2-\omega^2}},
\end{equation}
where $\omega_\mathrm{p}$ is a model parameter.
The dispersion equation $k=n_\mathrm{p}(\omega)\omega/c$ then takes the quadratic form \cite{Milonni2010}
\begin{equation}
 \omega^4-(\omega_\mathrm{p}^2+\omega_\mathrm{r}^2+k^2c^2)\omega^2+k^2c^2\omega_\mathrm{r}^2=0.
 \label{eq:nonlineardispersioneq}
\end{equation}
For each $k$, there are two positive solutions.
These solutions are given by \cite{Milonni2010}
\begin{equation}
 \omega_\pm\!=\!\sqrt{\frac{\omega_\mathrm{r}^2+\omega_\mathrm{p}^2+k^2c^2
\pm\sqrt{(\omega_\mathrm{r}^2+\omega_\mathrm{p}^2+k^2c^2)^2-4k^2c^2\omega_\mathrm{r}^2}}{2}}.
\end{equation}
These are called the upper $(+)$ and lower $(-)$ polariton branches and they
have been illustrated in Fig.~\ref{fig:branches}.
For the wave number $k=n_\mathrm{p}(\omega_0)\omega_0/c$ corresponding to
$\omega_0$ with $\omega_0>\omega_\mathrm{r}$, we obtain
$\omega_+=\omega_0$ and $\omega_-=n_\mathrm{p}(\omega_0)\omega_\mathrm{r}$ and with
$\omega_0<\omega_\mathrm{r}$ we obtain
$\omega_+=n_\mathrm{p}(\omega_0)\omega_\mathrm{r}$ and $\omega_-=\omega_0$.
We must restrict to the solution for which $\omega_i=\omega_0$.
The other solution of the dispersion equation has the same
wave number, but the frequency is very different from $\omega_0$.

The main difference between linear and nonlinear dispersion is that,
in the nonlinear case, the envelope of a wave packet becomes
distorted as it travels. This follows from the fact that the group velocity
$v_\mathrm{g}=\partial\omega(k)/\partial k$ is not constant but
different for different values of $k$. Therefore, the wave-number components
move at different velocities distorting the envelope of the wave packet.
However, if the wave packet has a range of frequencies that is narrow
enough compared to the nonlinearity, then $\omega(k)$ is necessarily
approximately linear over that narrow range and
the pulse distortion is small. Therefore, in the limit
of narrow frequency range, using the linear dispersion
relation in Eq.~\eqref{eq:lineardispersion} becomes an
accurate approximation.

\section{\label{sec:ocdmodel}Optoelastic continuum dynamics}

\subsection{Optical force density}
In previous literature, there has been extensive discussion on the appropriate
form of the force density acting on the medium under the
influence of time-dependent electromagnetic field \cite{Milonni2010}.
We have recently shown \cite{Partanen2017c} that there is only
one form of optical force density that is fully consistent
with the MP quasiparticle model and the underlying principles
of the special theory of relativity in the case of nondispersive
media \cite{Partanen2017c}. We generalize
this optical force density for dispersive media by writing
\begin{equation}
 \mathbf{f}_\mathrm{opt}(\mathbf{r},t)=-\varepsilon_0n_\mathrm{g}\mathbf{E}^2\nabla n_\mathrm{p}+\frac{n_\mathrm{p}n_\mathrm{g}-1}{c^2}\frac{\partial}{\partial t}\mathbf{E}\times\mathbf{H},
 \label{eq:opticalforcedensity}
\end{equation}
where $\mathbf{E}\times\mathbf{H}=\mathbf{S}$ is the instantaneous Poynting vector.
The expression of the optical force density given in Eq.~\eqref{eq:opticalforcedensity}
can be at this stage taken as a postulate. We will later on justify
it by showing that it is the only form of the optical force density that
enables covariant description of the light pulse, fulfills the conservation
law of momentum, and is also consistent with the MP quasiparticle model.
We have not found this specific form of the optical force density
in previous works.

In calculating the optoelastic force field,
we assume that the damping of the electromagnetic field due to
the transfer of field energy to the kinetic and elastic energies
of the medium by the optical force is negligible.
Adopting this perturbative
approach is justified as the effect of the fields on the dynamical state of
the material is such that the back action of the dynamics of the medium
on the state of the fields is extremely small.
The accuracy of this
approximation is estimated in Ref.~\cite{Partanen2017c} and
the conclusions are valid also for dispersive media
if there is no direct optical absorption related, e.g.,
to the electronic excitation of the medium.

\subsection{Newton's equation of motion}
In the OCD model, the coupling between the field and matter
is described by Newton's equation of motion.
As the atomic velocities are nonrelativistic,
Newton's equation of motion for the mass density
of the medium $\rho_\mathrm{a}(\mathbf{r},t)$ is given by
\begin{equation}
 \rho_\mathrm{a}(\mathbf{r},t)\frac{d^2\mathbf{r}_\mathrm{a}(\mathbf{r},t)}{dt^2}=\mathbf{f}_\mathrm{opt}(\mathbf{r},t)+\mathbf{f}_\mathrm{el}(\mathbf{r},t),
 \label{eq:mediumnewton}
\end{equation}
where $\mathbf{r}_\mathrm{a}(\mathbf{r},t)$ is the position- and time-dependent atomic
displacement field of the medium, $\mathbf{f}_\mathrm{opt}(\mathbf{r},t)$
is the optical force density experienced by atoms, given in Eq.~\eqref{eq:opticalforcedensity}, and
$\mathbf{f}_\mathrm{el}(\mathbf{r},t)$ is the elastic force
density between atoms that are displaced from
their initial equilibrium positions by the optical force density.

Close to equilibrium, the elastic forces between atoms are known to be well described by Hooke's law.
In the simple case of a homogeneous isotropic elastic medium, the elastic force density
in terms of the material displacement field $\mathbf{r}_\mathrm{a}(\mathbf{r},t)$
is well known to be given by \cite{Bedford1994}
\begin{equation}
 \mathbf{f}_\mathrm{el}(\mathbf{r},t)=\textstyle(B+\frac{4}{3}G)\nabla[\nabla\cdot\mathbf{r}_\mathrm{a}(\mathbf{r},t)]-G\nabla\times[\nabla\times\mathbf{r}_\mathrm{a}(\mathbf{r},t)],
 \label{eq:elasticforcedensity}
\end{equation}
where $B$ is the bulk modulus and $G$ is the shear modulus
of the medium \cite{Mavko2003}.
In more general anisotropic cubic crystals, Eq.~\eqref{eq:elasticforcedensity}
must be replaced with a more general form
given by the following set of componentwise equations \cite{Kittel2005}:
\begin{align}
 f_{\mathrm{el},x} =\;& C_{11}\frac{\partial^2r_{\mathrm{a},x}}{\partial x^2}
 +C_{44}\Big(\frac{\partial^2r_{\mathrm{a},x}}{\partial y^2}+\frac{\partial^2r_{\mathrm{a},x}}{\partial z^2}\Big)\nonumber\\
 &+(C_{12}+C_{44})\Big(\frac{\partial^2r_{\mathrm{a},y}}{\partial x\partial y}+\frac{\partial^2r_{\mathrm{a},z}}{\partial x\partial z}\Big),
 \label{eq:anisotropicforcex}
\end{align}
\begin{align}
 f_{\mathrm{el},y} =\;& C_{11}\frac{\partial^2r_{\mathrm{a},y}}{\partial y^2}
 +C_{44}\Big(\frac{\partial^2r_{\mathrm{a},y}}{\partial x^2}+\frac{\partial^2r_{\mathrm{a},y}}{\partial z^2}\Big)\nonumber\\
 &+(C_{12}+C_{44})\Big(\frac{\partial^2r_{\mathrm{a},x}}{\partial x\partial y}+\frac{\partial^2r_{\mathrm{a},z}}{\partial y\partial z}\Big),
 \label{eq:anisotropicforcey}
\end{align}
\begin{align}
 f_{\mathrm{el},z} =\;& C_{11}\frac{\partial^2r_{\mathrm{a},z}}{\partial z^2}
 +C_{44}\Big(\frac{\partial^2r_{\mathrm{a},z}}{\partial x^2}+\frac{\partial^2r_{\mathrm{a},z}}{\partial y^2}\Big)\nonumber\\
 &+(C_{12}+C_{44})\Big(\frac{\partial^2r_{\mathrm{a},x}}{\partial x\partial z}+\frac{\partial^2r_{\mathrm{a},y}}{\partial y\partial z}\Big),
 \label{eq:anisotropicforcez}
\end{align}
where $C_{11}$, $C_{12}$, and $C_{44}$ are elastic constants.
The forces given in Eqs.~\eqref{eq:anisotropicforcex}--\eqref{eq:anisotropicforcez}
simplify to the case of an isotropic medium in Eq.~\eqref{eq:elasticforcedensity}
by substitutions $C_{11}=B+\frac{4}{3}G$, $C_{12}=B-\frac{2}{3}G$, and $C_{44}=G$.

\subsection{Energy and momentum of the MP}
For a monochromatic field with angular frequency $\omega_0$ in a lossless dispersive medium,
the energy and momentum are given by \cite{Landau1984}
\begin{equation}
 E_\text{field}=\int\frac{1}{2}\Big[\frac{d(\varepsilon\omega_0)}{d\omega_0}\mathbf{E}^2+\frac{d(\mu\omega_0)}{d\omega_0}\mathbf{H}^2\Big]d^3r,
 \label{eq:fieldenergy}
\end{equation}
\begin{equation}
 \mathbf{p}_\text{field}=\int\frac{1}{c^2}\mathbf{E}\times\mathbf{H}d^3r.
 \label{eq:fieldmomentum}
\end{equation}
The momentum density of the field in Eq.~\eqref{eq:fieldmomentum} is
essentially of the Abraham form.
The momentum density in Eq.~\eqref{eq:fieldmomentum}
is also justified by the MP quasiparticle model as described below.

The energy density in the integrand of Eq.~\eqref{eq:fieldenergy} is known to be accurate only
in the limit of a monochromatic field. Here, we use it as an approximation
for light pulses. A more accurate but also more complicated expression for the energy density
of a finite light pulse in a dispersive medium is given in Ref.~\cite{Philbin2011}.

In the same way as done for a nondispersive medium in Ref.~\cite{Partanen2017c},
it can be easily shown that, in the OCD model, the energy and momentum of the MDW atoms are
given by
\begin{equation}
 E_\text{\tiny MDW}=\int\rho_\text{\tiny MDW}c^2d^3r
 \approx(n_\mathrm{p}n_\mathrm{g}-1)E_\text{field},
 \label{eq:mdwenergy}
\end{equation}
\begin{equation}
 \mathbf{p}_\text{\tiny MDW}=\int\rho_\mathrm{a}\mathbf{v}_\mathrm{a}d^3r
=\int\rho_\text{\tiny MDW}\mathbf{v}_\mathrm{g}d^3r
\approx(n_\mathrm{p}n_\mathrm{g}-1)\mathbf{p}_\text{field}.
 \label{eq:mdwmomentum}
\end{equation}
Here $\mathbf{v}_\mathrm{a}=d\mathbf{r}_\mathrm{a}/dt$ is the velocity
of atoms, $\mathbf{v}_\mathrm{g}$ is the group velocity vector, and the MDW mass density $\rho_\text{\tiny MDW}$ is given by
$\rho_\text{\tiny MDW}=\rho_\mathrm{a}-\rho_0$,
in which $\rho_0$ is the equilibrium mass density of the medium.
Thus, the MDW mass density corresponds to the excess mass density in the medium.
The total energy and momentum of the MP are given as sums
$E_\text{\tiny MP}=E_\text{\tiny MDW}+E_\text{field}$ and
$\mathbf{p}_\text{\tiny MP} =\mathbf{p}_\text{\tiny MDW}+\mathbf{p}_\text{field}$
resulting in
\begin{equation}
 E_\text{\tiny MP} \!=\!\!\int\!\Big\{\rho_\text{\tiny MDW}c^2+\frac{1}{2}\Big[\frac{d(\varepsilon\omega_0)}{d\omega_0}\mathbf{E}^2+\frac{d(\mu\omega_0)}{d\omega_0}\mathbf{H}^2\Big]\Big\}d^3r,
 \label{eq:mpenergy}
\end{equation}
\begin{equation}
 \mathbf{p}_\text{\tiny MP} 
 =\int\Big(\rho_\mathrm{a}\mathbf{v}_\mathrm{a}+\frac{1}{c^2}\mathbf{E}\times\mathbf{H}\Big)d^3r.
 \label{eq:mpmomentum}
\end{equation}

Following Appendix  B of Ref.~\cite{Partanen2017c},
it is also straightforward to present the energy and momentum
densities in the integrands of Eqs.~\eqref{eq:fieldenergy}--\eqref{eq:mpmomentum}
using the energy-momentum tensor formalism. The total
energy-momentum tensor of the MP can also be correspondingly split into parts
related to the electromagnetic field and the MDW.

\section{\label{sec:quasiparticlemodel}Mass-polariton quasiparticle model}

In the following, we generalize the MP quasiparticle model of Ref.~\cite{Partanen2017c}
for dispersive media. To emphasize the role of the MP as an intrinsic covariant
state of a single photon coupled to the medium, we neglect for the moment the
possible interface effects that occur when the photon enters the medium
and instead \emph{assume} that a photon having a field energy $\hbar\omega_0$
is propagating inside the medium.
Generalization of the present
work for full quantum optical description of the MP is left
for future works.

Instead, we use an analogy of
a single MP state to a very narrow wave packet in phase
space having a central frequency $\omega_0$ and field energy $\hbar\omega_0$.
In the OCD theory, such a wave packet can be made
arbitrarily close to a monochromatic wave.
Monochromatic components of such a wave packet propagate
at the phase velocity $v_\mathrm{p}$. \emph{First}, we assume that the field energy of the wave
packet will vanish in the frame propagating with velocity $v_\mathrm{p}$ ($F$ frame).
\emph{Second}, we assume that the frame moving with the group
velocity $v_\mathrm{g}$ ($R$ frame) is the rest frame of the MP
and accordingly the total momentum of the MP becomes zero in this frame.
\emph{Third}, in analogy with the case
of a nondispersive medium, we know that the kinetic energy of the atomic MDW
is extremely small in the laboratory frame ($L$ frame),
which is the initial rest frame of the medium.
As described in Ref.~\cite{Partanen2017c},
the mass $\delta m$ of the MDW is carried by atoms. Since the total
mass of atoms in the MDW is vastly larger than the mass
$\delta m$ carried by the MDW, the speed of atoms is
very small and, in particular, their kinetic energy is
extremely small in comparison with $\hbar\omega_0$.
Next we determine the total energy and momentum of the MP
by requiring that their values in the $L$ frame, $F$ frame, and $R$ frame
are related by the Lorentz transformation.

\emph{L frame}.
The total energy of the MP in $L$ frame is given by
$E_\text{\tiny MP}=\hbar\omega_0+\delta mc^2$.
The first term is the assumed fixed field energy.
The second term $\delta m$ is the mass energy carried by the MDW.
Note that, as discussed above, the kinetic energy of the MDW
is negligible in the $L$ frame.
The problem to be solved is to determine $\delta m$
and the total momentum $p_\text{\tiny MP}$ of the MP.

\emph{Lorentz transformation}.
When the $L$ frame energy and momentum of the MP
are transformed to any frame moving with constant velocity $v$
with respect to the $L$ frame, their values in the moving frame are given by
the Lorentz transformation as
\begin{equation}
 E_\text{\tiny MP}'=\gamma_v(E_\text{\tiny MP}-vp_\text{\tiny MP})=\gamma_v(\hbar\omega_0-vp_\text{\tiny MP})+\gamma_v\delta mc^2,
 \label{eq:LorentzE}
\end{equation}
\begin{equation}
 p_\text{\tiny MP}'=\gamma_v\Big(p_\text{\tiny MP}-\frac{vE_\text{\tiny MP}}{c^2}\Big)=\gamma_v\Big(p_\text{\tiny MP}-\frac{v\hbar\omega_0}{c^2}-v\delta m\Big),
 \label{eq:Lorentzp}
\end{equation}
where $\gamma_v=1/\sqrt{1-v^2/c^2}$ is the Lorentz factor.
In the following, we will show that $p_\text{\tiny MP}$ and $\delta m$
can be determined by investigating Eqs.~\eqref{eq:LorentzE} and \eqref{eq:Lorentzp}
in two special inertial frames: the $F$ frame and the $R$ frame.

\emph{F frame}.
First, we observe that, in Eq.~\eqref{eq:LorentzE}, the last term on the right
represents the transformed mass energy of the MP, while the first
term $\hbar\omega_0'=\gamma_v(\hbar\omega_0-vp_\text{\tiny MP})$
has its origin entirely in the field energy.
In the special case of the $F$ frame,
which propagates with the phase velocity $v=v_\mathrm{p}=c/n_\mathrm{p}$,
the frequency and the related field energy become zero as $\hbar\omega_0'\rightarrow 0$.
Therefore, we obtain
\begin{equation}
 p_\text{\tiny MP}=\frac{n_\mathrm{p}\hbar\omega_0}{c},
 \label{eq:pmp}
\end{equation}
which is of the Minkowski form as commonly defined
in literature for a dispersive medium \cite{Barnett2010b,Barnett2010a}.
Note that in the literature there exists also another rather commonly defined
form of the Minkowski momentum given by $p_\mathrm{M}=n_\mathrm{p}^2\hbar\omega_0/(n_\mathrm{g}c)$
\cite{Garrison2004}.
These momenta are discussed in more detail in Sec.~\ref{sec:comparisonwithprevious}.

\emph{R frame}.
Second, we consider the special case of the $R$ frame in which the total
momentum of the MP is zero by definition.
Inserting the momentum $p_\text{\tiny MP}$ from
Eq.~\eqref{eq:pmp} into Eq.~\eqref{eq:Lorentzp} and setting
$v=v_\mathrm{g}=c/n_\mathrm{g}$ and $p_\text{\tiny MP}'=0$,
we obtain
\begin{equation}
 \delta m =(n_\mathrm{p}n_\mathrm{g}-1)\hbar\omega_0/c^2.
 \label{eq:dm}
\end{equation}

As the final outcome, we have obtained unique values
for $p_\text{\tiny MP}$ and $\delta m$, given in Eqs.~\eqref{eq:pmp}
and \eqref{eq:dm}. With the Lorentz transformation in
Eqs.~\eqref{eq:LorentzE} and \eqref{eq:Lorentzp}, these values
can be used to unambiguously calculate the total energy and
momentum of the MP in arbitrary inertial frames.
Therefore, the MP quasiparticle is the only
model of a light quantum in a medium
that fully satisfies the Lorentz transformation
and is consistent with the phase and group velocities.

According to the special theory of relativity, we can write the
total energy of the MP in the $R$ frame as $m_0c^2$, where $m_0$ is the
rest mass of the structural system of the MP. Therefore, inserting
$p_\text{\tiny MP}$ and $\delta m$ from Eqs.~\eqref{eq:pmp} and \eqref{eq:dm}
into Eq.~\eqref{eq:LorentzE} together with $v=v_\mathrm{g}=c/n_\mathrm{g}$, we obtain in the $R$ frame
\begin{equation}
 m_0=n_\mathrm{p}\sqrt{n_\mathrm{g}^2-1}\,\hbar\omega_0/c^2.
 \label{eq:m0}
\end{equation}

The corresponding MP energy and momentum in the $L$ frame are then given by
\begin{align}
 E_\text{\tiny MP} &=\gamma_{v_\mathrm{g}}m_0c^2=n_\mathrm{p}n_\mathrm{g}\hbar\omega_0,\nonumber\\
 p_\text{\tiny MP} &=\gamma_{v_\mathrm{g}}m_0v_\mathrm{g}=\frac{n_\mathrm{p}\hbar\omega_0}{c}.
 \label{eq:energymomentumresults}
\end{align}
These results essentially generalize the results of Ref.~\cite{Partanen2017c}
for dispersive media. 
The energy and momentum in Eq.~\eqref{eq:energymomentumresults}
and the rest mass in Eq.~\eqref{eq:m0} fulfill the covariance condition
$E_\text{\tiny MP}^2-(p_\text{\tiny MP}c)^2=(m_0c^2)^2$.
Although knowing $\delta m$ is enough to
understand the mass transfer associated with the MP, $m_0$ is useful
for transparent understanding of the covariant MP state
of light in a medium.

Using the covariant energy-momentum ratio $E/p=c^2/v_\mathrm{g}$,
we can split the total MP momentum in Eq.~\eqref{eq:energymomentumresults}
into parts corresponding to the electromagnetic energy $E_\mathrm{field}=\hbar\omega_0$
and the MDW energy $E_\text{\tiny MDW}=\delta mc^2$. As a result, we obtain the field's
and MDW's shares of the total MP momentaum in the $L$ frame as
\begin{align}
 p_\text{\tiny MDW} &=\delta mv_\mathrm{g}=\Big(n_\mathrm{p}-\frac{1}{n_\mathrm{g}}\Big)\frac{\hbar\omega_0}{c},\nonumber\\
 p_\mathrm{field} &=p_\text{\tiny MP}-p_\text{\tiny MDW}=\frac{\hbar\omega_0v_\mathrm{g}}{c^2}=\frac{\hbar\omega_0}{n_\mathrm{g}c}.
 \label{eq:momentumsplitting}
\end{align}
The field's share of the momentum is of the Abraham form
and the MDW's share of the momentum corresponds to the difference
of the Minkowski and Abraham momenta.

Using Eqs.~\eqref{eq:pmp}--\eqref{eq:energymomentumresults} one
can easily show that the constant center of energy velocity (CEV) law,
essentially equal to Newton's first law, is fulfilled by the MP
theory also in the case of dispersive media. We apply the
conservation of momentum at the interface where the photon
enters a medium block. The photon momentum in vacuum must then be equal
to the sum of the MP momentum and the possible recoil momentum
received by a thin interface layer of the medium block.
We can write the momentum conservation law as
$\hbar\omega_0/c=p_\text{\tiny MP}+M_\mathrm{r}V_\mathrm{r}$,
where $M_\mathrm{r}=M-\delta m$ is the mass of the medium
block from which the mass transferred by the MP has been subtracted.
The center of energy velocity $V_\mathrm{r}$ of $M_\mathrm{r}$
can then be solved from the momentum conservation law as
$V_\mathrm{r}=(1-n_\mathrm{p})\hbar\omega_0/(M_\mathrm{r}c)$,
where we have used the transferred mass given in Eq.~\eqref{eq:dm}.
Writing the energy of the MP using its rest mass given in Eq.~\eqref{eq:m0} and observing that
the atomic velocities are certainly nonrelativistic, we can write the CEV law
before and after the photon has entered the medium as
\begin{equation}
 V_\mathrm{CEV}=\frac{\sum_iE_iv_i}{\sum_iE_i}=\frac{\hbar\omega_0 c}{\hbar\omega_0+Mc^2}=\frac{\gamma m_0c^2v_\mathrm{g}+M_\mathrm{r}c^2V_\mathrm{r}}{\gamma m_0c^2+M_\mathrm{r}c^2}.
 \label{eq:uniformmotion2}
\end{equation}
Here the summation is over all material particles and field quanta
and $E_i$ and $v_i$ are their energies and velocities.
The equality of the numerators divided by $c^2$
corresponds to the momentum conservation and
the equality of the denominators is nothing but
the conservation of energy.
Therefore, Eq.~\eqref{eq:uniformmotion2} directly shows that
the MP model obeys the constant CEV motion in dispersive media,
thus generalizing the result derived for nondispersive media
in Ref.~\cite{Partanen2017c}. This also explains
why the derivations of the Minkowski momentum assuming
zero rest mass for the light quantum in a medium lead
to violation of the constant CEV motion \cite{Barnett2010b,Barnett2010a}.

As a side product, by substituting $p_\text{\tiny MP}$ from Eq.~\eqref{eq:pmp} into
the first term of Eq.~\eqref{eq:LorentzE}, $\hbar\omega_0'=\gamma_v(\hbar\omega_0-vp_\text{\tiny MP})$, we obtain
$\hbar\omega_0'=\gamma_v(1-n_\mathrm{p}v/c)\hbar\omega_0$, which
is the well-known Doppler-shifted energy of a photon
in a medium in an arbitrary frame moving with the velocity $v$ with respect
to the $L$ frame \cite{ChenJi2011}.
Thus, the total MP momentum of the Minkowski form
in Eq.~\eqref{eq:pmp} can also be derived from the Doppler shift
\cite{Barnett2010b,Milonni2005}, which, however, must be used
as a part of the Lorentz transformation in Eqs.~\eqref{eq:LorentzE} and \eqref{eq:Lorentzp}
in order to enable the determination of the transferred mass $\delta m$ of the MP.

\vspace{-0.1cm}

\section{\label{sec:simulations}OCD simulations}

Above, we have derived the MP theory for dispersive media
using the complementary MP quasiparticle and OCD models.
In order to show the correspondence between these models and to illustrate the MDW and the
actual atomic displacements due to optoelastic forces in dispersive media,
we present numerical OCD simulations of a Gaussian light pulse
propagating in linearly and nonlinearly dispersive sample media.
To facilitate the planning of possible experiments,
we also compute the atomic displacements due to the MDW in silicon.

\subsection{\label{sec:simulationsvisualization}Visualization of the node structure of the MDW}

\begin{figure*}
\centering
\includegraphics[width=\textwidth]{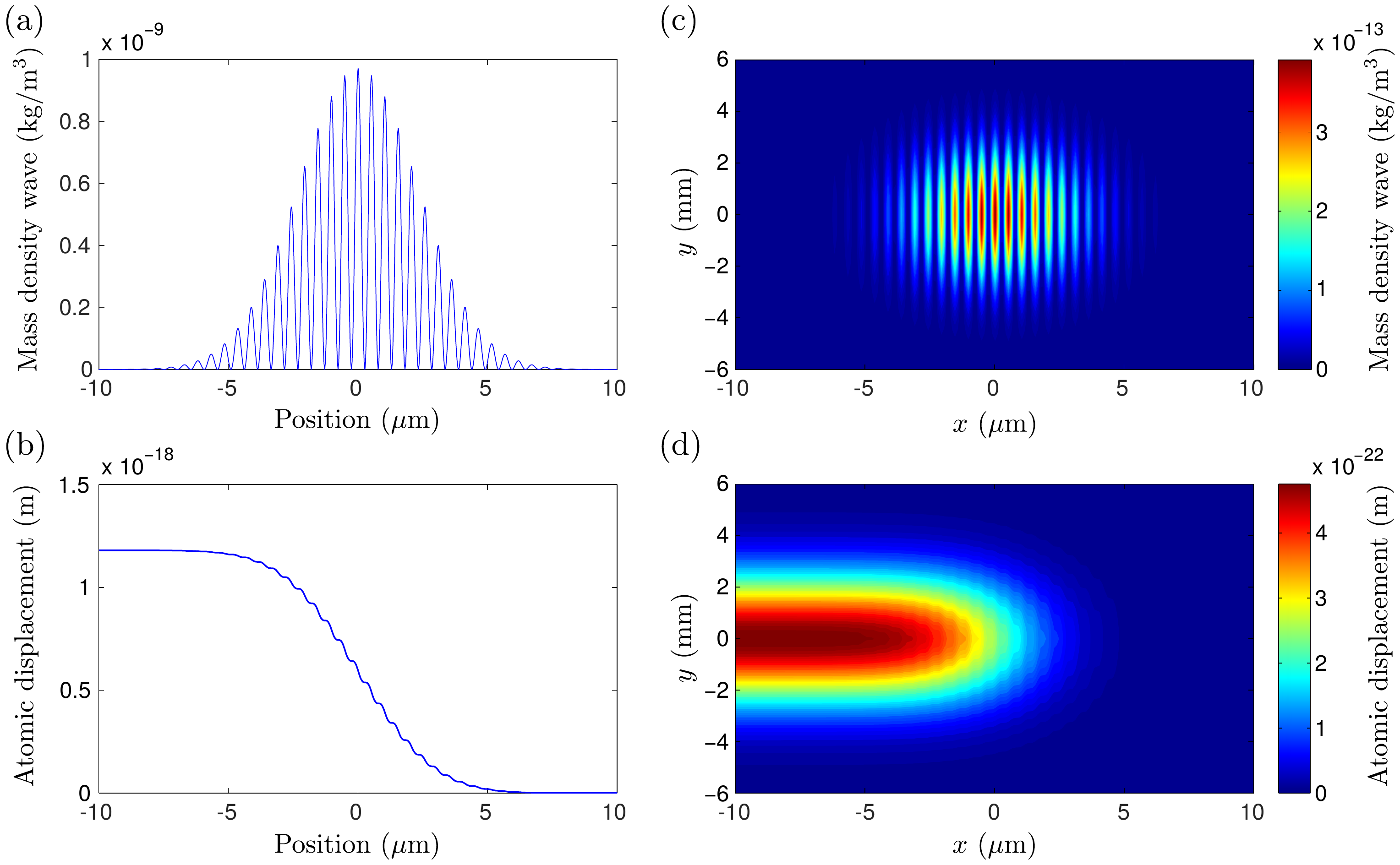}
\caption{\label{fig:linear}
(Color online) Illustration of (a) the MDW and (b) the atomic displacements
in a linearly dispersive material
where the phase and group refractive indices for the central frequency
are $n_\mathrm{p}=1.5$ and $n_\mathrm{g}=2$.
The Gaussian light pulse has a vacuum wavelength $\lambda_0=1550$ nm, $\Delta t_\mathrm{FWHM}=27$ fs,
and energy $U_0=1$ $\mu$J per cross-sectional area of diameter $d=100$ $\mu$m.
The pulse parameters are close to technological
feasibility limit but they are chosen to visualize the node structure of the MDW.
Panel (c) shows the MDW and (d) shows the atomic displacements of
a three-dimensional light pulse with finite lateral dimensions
in the plane $z=0$ m.}
\vspace{-0.2cm}
\end{figure*}

\subsubsection{Linearly dispersive medium}

First, we apply the OCD model to illustrate the node structure of the
MDW and the actual atomic displacements
due to a Gaussian light pulse in a linearly dispersive material.
The Gaussian light pulse of Eq.~\eqref{eq:electricfield}
is assumed to have a vacuum wavelength of $\lambda_0=1550$ nm and
a total electromagnetic energy of $U_0=1$ $\mu$J.
We assume that the relative spectral width
of the pulse, in our example, is $\Delta\omega/\omega_0=\Delta k_0/k_0=0.05$
corresponding to $\Delta t_\mathrm{FWHM}=27$ fs.
The FWHM is fixed to this close to feasibility limit value
to make the node structure of the MDW visible.
In our simulations, we use space discretization of
$h_x=\lambda/40$ and time discretization of $h_t=2\pi/(40\omega_0)$
that are sufficiently dense compared to the scale of the harmonic cycle.
The computational details of the simulation are described in
Appendix C of Ref.~\cite{Partanen2017c}.

For visualization needs, we use here an artificial example material
for which the refractive indices at the central frequency $\omega_0=2\pi c/\lambda_0$
are $n_\mathrm{p}=1.5$ and $n_\mathrm{g}=2$.
The chosen phase refractive index is close to typical values
for glasses but we have made the group refractive index somewhat
larger to enable the visual separation of the phase velocity dynamics
of the nodes inside the Gaussian envelope.
The diameter $d$ of the circular cross-sectional area $A=\pi(d/2)^2$
is assumed to be $d=100$ $\mu$m,
which is chosen to be large enough so that the
resulting maximum value $3.3\times10^{11}$ W/cm$^2$
of the Poynting vector averaged over the harmonic cycle is below the
bulk value of the breakdown threshold irradiance of many common materials, e.g.,
$5.0\times10^{11}$ reported for fused silica \cite{Smith2007}.
The equilibrium density of the material is assumed to be $\rho_0=2400$ kg/m$^3$,
the material is assumed to be isotropic,
and we use the value $B=50$ GPa for the bulk modulus and $G=25$ GPa
for the shear modulus. These values are close to typical values of
the corresponding quantities for glass.

Figure \ref{fig:linear}(a) shows the simulated MDW as a function of position
when the light pulse is propagating at the position $x=0$ $\mu$m.
The time-dependent simulation is presented
as a video file in the Supplemental Material \cite{supplementaryvideo}.
The MDW equals the difference of the disturbed mass
density $\rho_\mathrm{a}(\mathbf{r},t)$
and the equilibrium mass density $\rho_0$
of the medium and it is obtained by solving
Newton's equation of motion in Eq.~\eqref{eq:mediumnewton}.
The MDW is driven by the optoelastic forces due to the Gaussian
light pulse. The envelope of the MDW clearly follows the Gaussian
form of the pulse as expected. As the light pulse
is not very long compared to the harmonic cycle,
the node structure of the MDW can be seen in the
same scale with the Gaussian envelope.
When we integrate the MDW mass density
in Fig.~\ref{fig:linear}(a), we obtain the total
transferred mass of $2.23\times 10^{-23}$ kg. Dividing this by
the photon number of the light pulse, we then obtain the value of
$1.60$ eV/$c^2$ for the transferred mass per photon.
Within the relative error of $10^{-4}$, this equals the MP
quasiparticle value obtained from Eq.~\eqref{eq:dm}.
For a more detailed discussion of the correspondence between
the MP quasiparticle and OCD approaches, see Sec.~\ref{sec:comparison}.

Figure \ref{fig:linear}(b) shows the atomic displacements corresponding
to MDW in Fig.~\ref{fig:linear}(a), again,
as a function of position.
On the left of the light pulse,
the atomic displacement has a constant value of
$r_\mathrm{a,max}=1.18\times10^{-18}$ m.
This follows from the optical force in the second term of Eq.~\eqref{eq:opticalforcedensity}.
Within the relative error of $10^{-4}$, we obtain
$r_\mathrm{a,max}=\delta M/(\rho_0 A)$, where $\delta M=N_0\delta m$
is the total transferred mass of the light pulse. The leading edge of the
optical pulse is propagating to the right approximately at the position
$x=7$ $\mu$m. Therefore, to the right of $x=7$ $\mu$m, the atomic displacement is zero.
The optoelastically driven MDW is manifested by the fact that atoms
are more densely spaced at the position of the light pulse as the atoms
on the left of the pulse have been displaced forward and the atoms
on the right of the pulse are still at their equilibrium positions.
The momentum of atoms in the MDW is obtained by 
integrating the classical momentum density as given in Eq.~\eqref{eq:mdwmomentum}
at an arbitrary time.

We also illustrate the MDW and the atomic displacements due to a three-dimensional
light pulse. This light pulse is only an approximative solution of
Maxwell's equations. It
is obtained from the one-dimensional pulse
described by Eq.~\eqref{eq:electricfield} by adding
additional $y$ and $z$ dependencies by using factors
$e^{-(\Delta k_y)^2y^2/2}$ and $e^{-(\Delta k_z)^2z^2/2}$.
As reasoned in Ref.~\cite{Partanen2017c}, this approximation
becomes accurate if $\Delta k_y$ and $\Delta k_z$ are
sufficiently small compared to the wave number of
the central frequency in the medium equal to $k_\mathrm{0,med}=n_\mathrm{p}k_0$.
In our example, we use $\Delta k_y=\Delta k_z=10^{-4}k_0$,
which are small so that the approximation is well justified for
our visualization purposes.

The contour plot in Fig.~\ref{fig:linear}(c) shows
the MDW of the three-dimensional
Gaussian pulse in the plane $z=0$ m. The corresponding time-dependent
simulation is presented as a video file in the Supplemental Material \cite{supplementaryvideo}.
The three-dimensional pulse
differs from the one-dimensional pulse
in Fig.~\ref{fig:linear}(a) by its finite lateral
dimensions as described above. The values of the MDW in Fig.~\ref{fig:linear}(c)
are thus smaller than the values in Fig.~\ref{fig:linear}(a) due
to the smaller value of the energy per cross-sectional area.

The contour plot in Fig.~\ref{fig:linear}(d) presents the $x$ component
of the atomic displacements due to the three-dimensional
Gaussian pulse in the plane $z=0$ m. The values of the atomic displacement in Fig.~\ref{fig:linear}(d)
are smaller than the values in the one-dimensional case in Fig.~\ref{fig:linear}(b), again, due
to the smaller value of the energy per cross-sectional area.

\subsubsection{Nonlinearly dispersive medium}

Next we investigate the MDW in a nonlinearly dispersive example material.
The nonlinear dispersion is described by the simple Lorentz model of
a dielectric in Sec.~\ref{sec:dispersionnonlinear}.
We use the same parameters for the Gaussian pulse as above.
The only difference is the use of the nonlinear dispersion relation
in Eq.~\eqref{eq:nonlineardispersioneq}, in which the model parameters
$\omega_\mathrm{r}=1.632\,99\omega_0$ and
$\omega_\mathrm{p}=1.443\,38\omega_0$ have been determined
so that the phase and group refractive indices have the same
values $n_\mathrm{p}=1.5$ and $n_\mathrm{g}=2$
for the central frequency of the pulse as above.
As $\omega_0<\omega_\mathrm{r}$, the dispersion relation corresponds to
the lower polariton branch in Fig.~\ref{fig:branches}.

\begin{figure}
\centering
\includegraphics[width=\columnwidth]{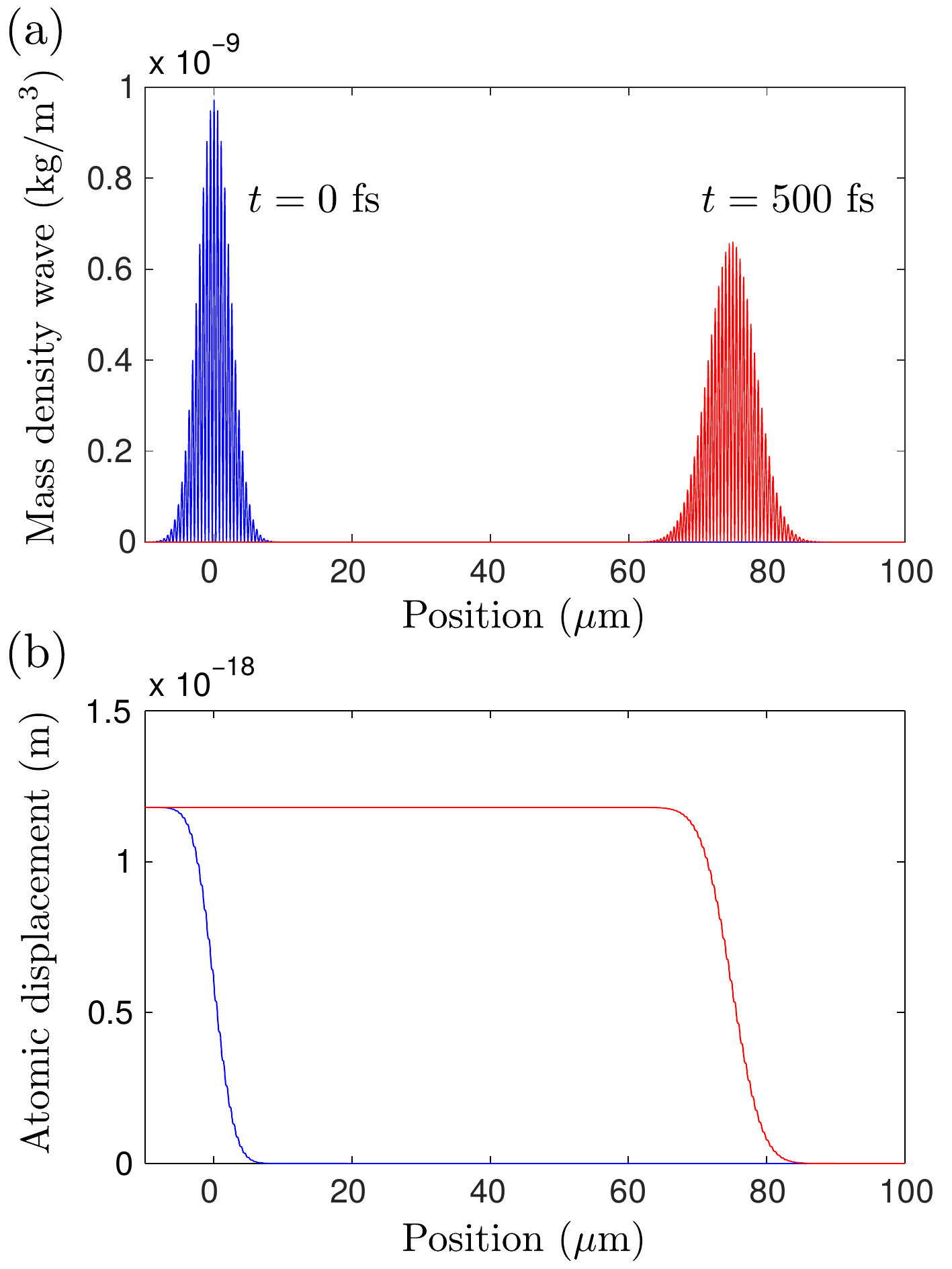}
\vspace{-0.5cm}
\caption{\label{fig:nonlinear}
(Color online) (a) The MDW and (b) the atomic displacement of a Gaussian
light pulse in a nonlinearly dispersive material as a function of the position
at two instances of time: $t=0$ fs (blue) and $t=500$ fs (red).
The vacuum wavelength is $\lambda_0=1550$ nm and the corresponding
phase and group refractive indices are $n_\mathrm{p}=1.5$ and $n_\mathrm{g}=2$. }
\vspace{-0.2cm}
\end{figure}

We start by briefly discussing how
the nonlinear dispersion relation affects the pulse shape
in comparison with the linear dispersion relation.
At $t=0$ fs, the electric field of the pulse is determined
by its Fourier components that are chosen to be of
the same Gaussian form as in the case of linear
dispersion above. In dispersive media, the Fourier
components of the magnetic field are given by 
$\tilde H(k)=Z(k)\tilde E(k)$, where the proportionality factor
$Z(k)=\sqrt{\mu[\omega(k)]/\varepsilon[\omega(k)]}$
is the $k$-dependent wave impedance.
Due to this $k$-dependent
proportionality factor, the Fourier components
of the magnetic field and the resulting pulse shapes
are modified depending on the dispersion relation even at $t=0$ fs.
However, for light pulses with a narrow spectral width, the
deviation in the pulse shape between the nonlinear and the corresponding linear dispersion relation
is typically very small for $t=0$ fs. At later times $t>0$ fs, the dispersion-modified
time dependence through the exponential factor $e^{-i\omega(k)t}$
of the fields in Eqs.~\eqref{eq:electricfieldgeneral}
and \eqref{eq:magneticfieldgeneral}
more clearly affects the pulse shape.
Therefore, in order to illustrate the effect of nonlinear dispersion,
we compare the MDW pulse shapes at $t=0$ and $t=500$ fs.

Figure \ref{fig:nonlinear}(a) presents the MDW of
the Gaussian pulse in the nonlinearly dispersive medium at these two instances of time.
The corresponding time-dependent
simulation is presented as a video file in the Supplemental Material \cite{supplementaryvideo}.
The pulse on the left in Fig.~\ref{fig:nonlinear}(a) corresponds to the pulse
at $t=0$ fs while the pulse on the right corresponds
to the pulse at $t=500$ fs. During this time interval,
the pulse has propagated to the right a distance
of $75$ $\mu$m. One can clearly see that the pulse
has become lower and broadened when compared
to the initial pulse. This effect follows purely from
the nonlinearity of the dispersion relation.
If the dispersion relation would be perfectly linear,
this broadening would not occur as the Gaussian
envelope would maintain its width.

Figure \ref{fig:nonlinear}(b) shows the atomic displacements
corresponding to the MDW in Fig.~\ref{fig:nonlinear}(a)
at $t=0$ and $t=500$ fs. The most clear physical
difference between the atomic displacements at these
two moments of time is that the slope of the atomic
displacement curve in Fig.~\ref{fig:nonlinear}(b)
is lowered at $t=500$ fs. This follows from the
broadening of the light pulse. The atomic displacements
on the left of the pulse are approximatively equal as expected.
This constant value, $r_\mathrm{a,max}$, depends on the phase and group
refractive indices and the density of the material, but it is only
slightly affected by the nonlinearity
of the dispersion relation. This can be seen by observing that
the constant value of the atomic displacement
on the left of the light pulse in Fig.~\ref{fig:nonlinear}(b)
is closely equal to the corresponding value in Fig.~\ref{fig:linear}(c).
This is related to the fact that we have used the same density
for the material and defined
the nonlinear dispersion relation so that the phase
and group refractive indices for the central frequency in our nonlinear case
are equal to the same quantities in the case of linear
dispersion above.

\vspace{-0.2cm}

\subsubsection{Continuous wave}

By changing the Gaussian light pulse to a top-hat pulse
and by making the length of the pulse very large,
we can also use the time-dependent OCD model to simulate
a continuous-wave (cw) laser beam.
The cw beam deserves a separate comment since
it has been extensively discussed in previous theoretical
works and in the analysis of experiments.
Previous theoretical works have often concluded that,
since the time average of the Abraham force given by the second
term of Eq.~\eqref{eq:opticalforcedensity} is zero
for the cw field, its effect is not directly observable
\cite{Milonni2010}. This conclusion is not sound.
The maxima of the cw field energy give rise to alternative
acceleration and deceleration of the atoms in the medium
in the direction of the light beam. As a net effect,
the atoms are displaced in the direction of the beam
and, in the average, they also carry momentum.
This changes the dynamical state of the medium
and also leads to shift of the atomic density which
must be accounted for in the analysis of experiments.
Note that in the simulations of the dynamical state
of the medium using the OCD method, whether we
analyze a light pulse in a solid using elasticity theory
or in a liquid using Navier-Stokes equation,
we cannot assume that the medium is incompressible,
which is often done \cite{Leonhardt2014}.
In a perfectly incompressible medium, the medium dynamics
cannot follow the time and position dependence
of the electromagnetic field in a way governed by
Newton's equation of motion in Eq.~\eqref{eq:mediumnewton}.

In long time scales,
the elastic forces that try to restore the mass equilibrium
in the medium also play an important role.
Assuming the geometry of a medium block whose transverse boundaries
are held fixed by external forces, the OCD model
leads to an equilibrium where the forward
mass transfer due to the MDW is balanced by
the backward mass transfer due to elastic waves.
The accumulation of the elastic waves together
with the absorption of photons also lead
to heating of the medium block. Therefore,
in order to obtain an equilibrium
in the simulation, one must also account
for the transport of the generated heat
over the boundaries of the medium block
by conduction and radiation. The detailed
study of this cw case is left as
a topic of further work.

\vspace{-0.2cm}

\subsection{Estimating atomic displacements of the MDW in silicon}

\vspace{-0.1cm}

\begin{figure}
\centering
\includegraphics[width=0.9\columnwidth]{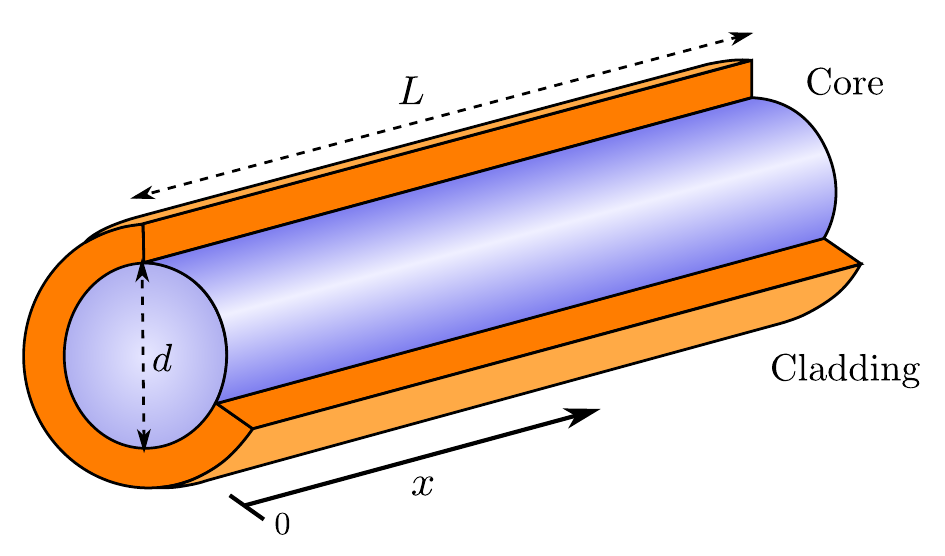}
\vspace{-0.2cm}
\caption{\label{fig:fiber}
(Color online) Schematic illustration of a waveguide or an optical fiber
with a core diameter $d$ and length $L$. The transferred mass of the MDW
is to be measured as the shift of atoms on the surface of the waveguide
at $x=L/2$ just after the light pulse has gone.}
\vspace{-0.2cm}
\end{figure}

Next we study how the atomic displacement of the MDW
depends on the pulse energy and the diameter of the
cross-sectional area. These calculations are presented
for designing experimental setups for the measurement
of the transferred mass of a light pulse.
We simulate a one-dimensional
Gaussian pulse in silicon for different pulse
energies, cross-sectional areas, and $\Delta t_\mathrm{FWHM}$.
The computed atomic displacements correspond to the
experimental arrangement in which the given
pulse energy is propagating in a waveguide or an optical fiber
as schematically illustrated in Fig.~\ref{fig:fiber}.
Due to the interface effects,
the cross-sectional area of the fiber cannot be directly compared
with the cross-sectional area of our calculations.
The core cross section of the waveguide or fiber should
be corrected for the possible cladding layer,
metallic coating, and other factors that influence
the spreading of the pulse energy in the transverse direction.
In detailed calculations,
the waveguide dispersion should also be taken into account.
All these factors can be easily accounted for in the OCD simulations.

The phase and group refractive indices of silicon are given by
$n_\mathrm{p}=3.4757$ and 
$n_\mathrm{g}=3.5997$ for $\lambda_0=1550$ nm \cite{Li1980}. The density
is $\rho_0=2329$ kg/m$^3$ \cite{Lide2004} and the elastic
constants in the direction of the (100) plane
are $C_{11}=165.7$ GPa, $C_{12}=63.9$ GPa, and $C_{44}=79.6$ GPa
\cite{Hopcroft2010}. These elastic constants correspond to the
bulk modulus of $B=(C_{11}+2C_{12})/3=97.8$ GPa and the shear modulus of $G=C_{44}=79.6$ GPa.

\begin{figure}
\centering
\includegraphics[width=\columnwidth]{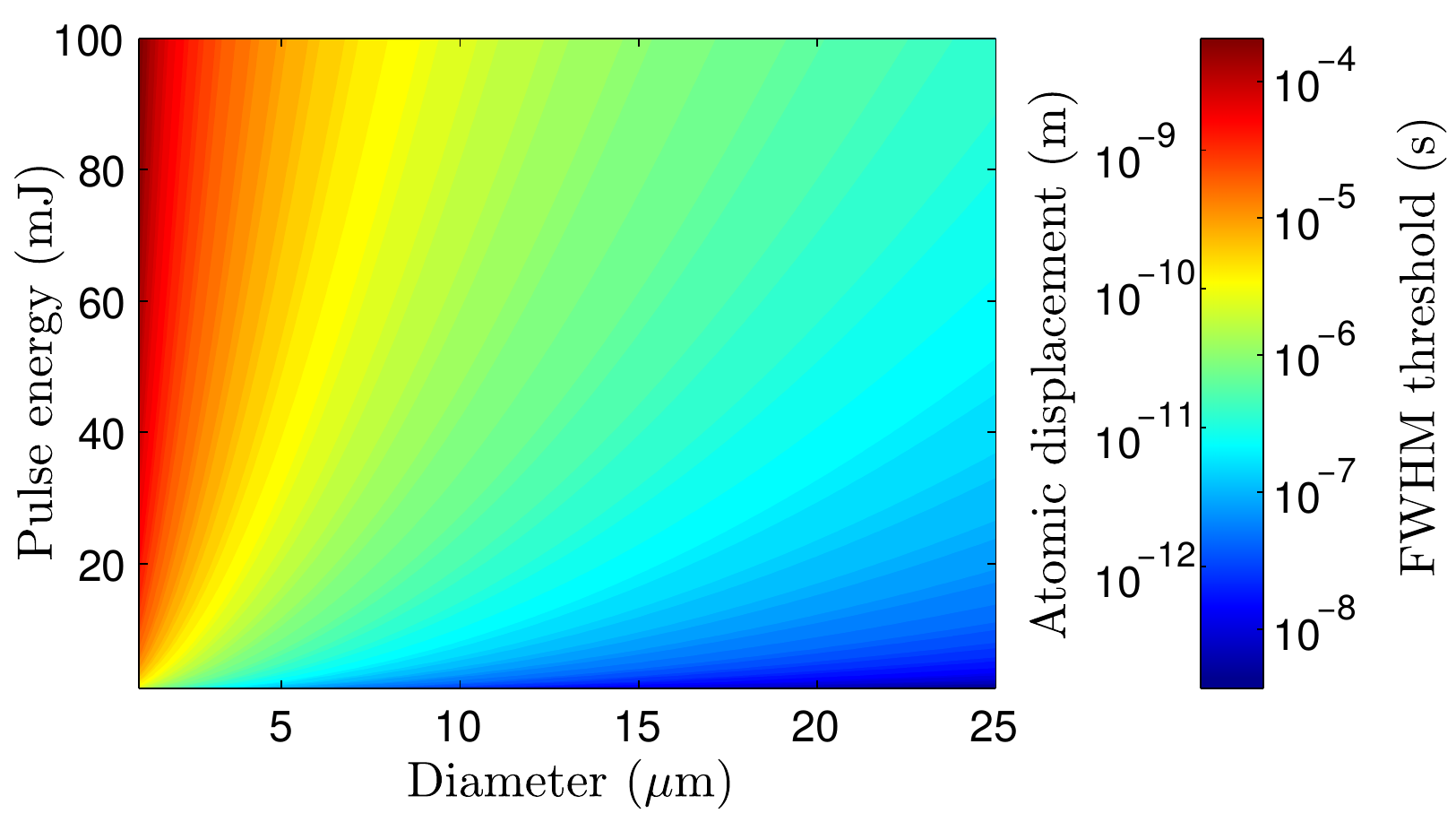}
\vspace{-0.4cm}
\caption{\label{fig:variation}
(Color online) The total atomic displacement of the MDW of a Gaussian
light pulse in silicon as a function of the pulse energy and the diameter of
the cross-sectional area of the pulse.
The vacuum wavelength is $\lambda_0=1550$ nm and the corresponding
phase and group refractive indices are $n_\mathrm{p}=3.4757$ and $n_\mathrm{g}=3.5997$. 
The second color-bar axis shows the threshold
$\Delta t_\mathrm{th}$ of the pulse obtained by requiring that the bulk value
of the breakdown threshold irradiance of the material is not exceeded.}
\vspace{-0.2cm}
\end{figure}

Figure \ref{fig:variation} shows the atomic displacement
as a function of the pulse energy and the diameter of the
cross-sectional area. Compared to the femtosecond
pulses above, we here assume longer pulses with
$\Delta t_\mathrm{FWHM}>1$ ns. Therefore, the correspondence of the
MP quasiparticle and the OCD models is very accurate
and we can use the quasiparticle model result
$r_\mathrm{a,max}=\delta M/(\rho_0A)$ for the maximum atomic displacement $r_\mathrm{a,max}$.
Using $\delta M=(n_\mathrm{p}n_\mathrm{g}-1)U_0/c^2$
and $A=\pi(d/2)^2$, where $d$ is the diameter of the cross-sectional area,
we then obtain $r_\mathrm{a,max}=(n_\mathrm{p}n_\mathrm{g}-1)U_0/[c^2\rho_0\pi(d/2)^2]$.
Hence the atomic displacement depends linearly on the
pulse energy while it is inversely proportional to
the cross-sectional area.
Consequently, in Fig.~\ref{fig:variation},
the atomic displacement is seen to be
large for high pulse energies and for small cross-sectional areas
as expected.

\vspace{-0.2cm}

\subsubsection{Influence of the material breakdown irradiance}

We also evaluate the minimum $\Delta t_\mathrm{FWHM}$ of a Gaussian pulse
that is needed to produce the corresponding atomic displacement without
exceeding the bulk value of the breakdown threshold
irradiance of the material. Using the total electromagnetic
energy of the pulse given by $U_0$,
and the cross-sectional area of the pulse given by
$A=\pi(d/2)^2$, this threshold
$\Delta t_\mathrm{FWHM}$, denoted by $\Delta t_\mathrm{th}$, is calculated as 
$\Delta t_\mathrm{th}=2U_0/[\pi(d/2)^2I_\mathrm{th}]$,
where $I_\mathrm{th}$ is the bulk value of the breakdown threshold
irradiance of the material. The corresponding
fluence is $F_\mathrm{th}=2U_0/[\pi(d/2)^2]$.
The factor $2$ comes from the fact that
the pulse is Gaussian and not a top-hat pulse with
constant irradiance. For silicon with $\lambda_0=1550$ nm,
the bulk value of the breakdown threshold energy
density has been reported to be $u_\mathrm{th}=13.3$ J/cm$^3$ \cite{Cowan2006},
which corresponds to the threshold irradiance of
$I_\mathrm{th}=u_\mathrm{th}v_\mathrm{g}=1.11\times 10^{11}$ W/cm$^2$.
These are values averaged over the harmonic cycle.

The threshold $\Delta t_\mathrm{th}$ of a Gaussian pulse calculated as
explained above is presented by the second
color-bar axis in Fig.~\ref{fig:variation}.
Using the relations above, the scaling between
the atomic displacement and the threshold $\Delta t_\mathrm{th}$
is given by $r_\mathrm{a,max}/\Delta t_\mathrm{th}=(n_\mathrm{p}n_\mathrm{g}-1)I_\mathrm{th}/(2c^2\rho_0)$.
This clearly indicates that, in order to obtain
large atomic displacements for a given pulse
energy, it is beneficial to have a material
with a high refractive index, high breakdown
threshold irradiance, and relatively small mass
density. In Fig.~\ref{fig:variation}, one can see
that, in order to obtain atomic displacements
larger than $1$ nm in silicon without breaking
the material, the pulse width must
be larger than $\Delta t_\mathrm{th}=33$ $\mu$s.

\subsubsection{Displacement of atoms due to optical absorption}

In measuring the atomic displacements due to the MDW, one essential point
is to ensure that the momentum transfer due to
optical absorption of the material
is not too large so that the resulting atomic movement
would exceed the photon mass drag effect.
Therefore, we estimate the atomic displacement
and the atomic velocity resulting from the optical absorption.
The mass of a cylindrical medium block with a diameter $d$
and length $L$,
or the core of the waveguide in Fig.~\ref{fig:fiber},
is given by $M=\rho_0\pi(d/2)^2L$.
The momentum absorbed by this medium block is given by
$P_\mathrm{abs}=(1-e^{-\alpha L})n_\mathrm{p}U_0/c\approx \alpha L n_\mathrm{p}U_0/c$,
where $\alpha$ is the small absorption coefficient of the medium.
The velocity obtained by the medium block is then
$V_\mathrm{abs}=P_\mathrm{abs}/M\approx \alpha n_\mathrm{p}U_0/[c\rho_0\pi(d/2)^2]$.
In the time scale of $\Delta t_\mathrm{FWHM}$, the resulting
atomic displacement is given by
$X_\mathrm{abs}=V_\mathrm{abs}\Delta t_\mathrm{FWHM}$.

In the case of silicon, absorption is very low at $\lambda_0=1550$ nm.
The measurements by Schinke \emph{et al.}~\cite{Schinke2015} and Green \cite{Green2008}
for $\lambda_0=1450$ nm give $\alpha\approx10^{-8}$ cm$^{-1}$ and the absorption
is known to decrease towards $\lambda_0=1550$ nm. Therefore, we can conservatively estimate
$\alpha=10^{-8}$ cm$^{-1}$. Using $\Delta t_\mathrm{FWHM}=\Delta t_\mathrm{th}=33$ $\mu$s and
$d=2.5$ $\mu$m corresponding to $r_\mathrm{a,max}=1.0$ nm atomic displacement due to the MDW,
and solving the threshold pulse energy
from $\Delta t_\mathrm{th}=2U_0/[\pi(d/2)^2I_\mathrm{th}]$,
we obtain $U_0=90$ mJ.
The velocity of atoms is then $V_\mathrm{abs}=9.1\times 10^{-8}$ m/s and,
in the time scale of $\Delta t_\mathrm{th}$, the resulting
atomic displacement is given by
$X_\mathrm{abs}=3.0$ pm. This atomic displacement due to optical absorption
is clearly smaller than $r_\mathrm{a,max}=1.0$ nm
following from the photon mass drag effect. Therefore,
optical absorption is not expected to prevent
measurements of the atomic displacements due to the photon
mass drag effect. This result strongly supports
the experimental feasibility of the measurement of
the transferred mass of the MDW.

We have also considered the thermal expansion
following from the optical absorption. Using the well-known
specific-heat capacity and thermal-expansion coefficients, it can
be shown that the thermal expansion does not lead to measurable
atomic displacements in the middle part of the fiber in the time
scale of $\Delta t_\mathrm{FWHM}$ that is shorter than the
time that it takes for sound waves to travel through the fiber.
This is also related to the longitudinal relaxation studied below.

\subsubsection{Transverse relaxation}

In the experimental verification of the transferred mass of the MDW,
one also has to account for the phonon relaxation of the atomic displacements due to the MDW.
This relaxation takes place at the velocity of sound and it is governed
by Eqs.~\eqref{eq:anisotropicforcex}--\eqref{eq:anisotropicforcez} of the OCD model.
The relaxation effect has been briefly studied in Ref.~\cite{Partanen2017c}.
If a three-dimensional light pulse propagates inside a medium or in the core of an optical fiber
that has a cladding, the MDW displaces atoms as shown
in Fig.~\ref{fig:linear}(d). The atoms along the path of the MP are 
displaced forward while the atoms in the surrounding layers are
not shifted. This results
in a shear strain field along the path of the MP.
The transverse relaxation refers to the relaxation of the strain field
so that atoms in the displaced region are shifted backwards and atoms
in the surrounding layers are shifted forwards. After the transverse
relaxation, the longitudinal strain becomes constant across the cross section
of the waveguide.

The relaxation of the strain field is quite fast in optical fibers where
the distances to be traveled by phonons in the transverse direction are very short.
Using the longitudinal velocity of sound in silicon,
given by $v_\parallel=\sqrt{C_{11}/\rho_0}=8435$ m/s, the time at which a sound wave
propagates, e.g., a distance of $1$ mm is $1.2$ ns. This is obviously very short
compared to the time scale of $\Delta t_\mathrm{FWHM}=33$ $\mu$s used above.
Thus, as a net effect, this transverse relaxation takes place in a time
scale that is shorter than the passing of the pulse and we can approximate
that, in a narrow waveguide, atoms are displaced in the longitudinal direction
by the same amount in the middle and at the surface of the waveguide.
After the transverse relaxation, the constant atomic displacement is reduced to
$r_\mathrm{a,relaxed}=r_\mathrm{a,max}\rho_0A/(\rho_\mathrm{eff}A_\mathrm{tot})$,
where $A_\mathrm{tot}$ is the total cross-sectional area of all layers and
$\rho_\mathrm{eff}$ is the effective mass density of the cross-sectional
area given by $\rho_\mathrm{eff}=\sum_i\rho_iA_i/A_\mathrm{tot}$, where
the sum is taken over all material layers and $\rho_i$ and $A_i$
are the densities and cross-sectional areas of the corresponding layers.

The time constant of the transverse relaxation is much
shorter than the pulse width $\Delta t_\mathrm{FWHM}$ for
structures where the atomic displacement is potentially measurable.
Therefore, it is advantageous to keep the waveguide diameter as small
as possible considering the effectivity of the coupling
of the light source to the waveguide and the technical
processing aspects of fabricating it. This suggests that the narrower
the waveguide is, the larger is the atomic displacement and
the breakdown of the material can be prevented by
increasing the pulse width $\Delta t_\mathrm{FWHM}$.
However, the longitudinal relaxation described in the next subsection
will set a limit for increasing $\Delta t_\mathrm{FWHM}$.

\subsubsection{Longitudinal relaxation}

After the transverse relaxation has taken place, further relaxation
can only occur starting from the ends of the fiber, which have experienced
recoil effects and which may be attached to some part of the experimental
setup that tries to keep them fixed. If the fiber is long enough,
these longitudinal relaxation waves starting from the ends of the fiber have
not time enough to reach the middle part of the fiber where
the atomic displacement is to be measured. The distance traveled
by sound in silicon in the time scale of $\Delta t_\mathrm{FWHM}=33$ $\mu$s
is $28$ cm. Therefore, a fiber with length $L>56$ cm is sufficient
to avoid the longitudinal relaxation from having an effect on the measured value
of the atomic displacement assuming that the atomic displacement in
the middle part of the fiber is measured just after the light pulse has gone.

\section{\label{sec:comparison}Discussion and comparison of the MP and OCD results}

\subsection{Dependence on the pulse width}

We have shown in Ref.~\cite{Partanen2017c} that, in the case of nondispersive media, the MP and OCD models give equal results within the relative numerical accuracy of the OCD simulations.
For a dispersive medium, the comparison of the MP and OCD momenta
becomes more subtle. The derivation of the MP model in Sec.~\ref{sec:quasiparticlemodel} assumes
infinitely narrow pulse in the frequency domain while the OCD model
involves integration over partial waves and thus accounts for the
frequency-dependent dispersion relation.
Using Eqs.~\eqref{eq:fieldenergy}--\eqref{eq:mpmomentum} and
\eqref{eq:pmp}--\eqref{eq:momentumsplitting}, the total transferred mass and the total momentum
and the momentum of the field and the
MDW can be written for a dispersive medium as given in Table \ref{tbl:table}.

\begin{table}
 \setlength{\tabcolsep}{0.0pt}
 \renewcommand{\arraystretch}{2.4}
 \caption{\label{tbl:table}
 The transferred mass, the total momentum, the field's share of the
 momentum, and the MDW's share of the momentum calculated
 by using the MP and OCD models. Here $N_0=U_0/\hbar\omega_0$
 is the photon number of the pulse.}
\begin{tabular}{ccc}
   \hline\hline
   & OCD & MP \\[4pt]
   \hline
   $\delta M$ & $\displaystyle\int\rho_\text{\tiny MDW}d^3r$ & $\displaystyle(n_\mathrm{p}n_\mathrm{g}-1)\frac{N_0\hbar\omega_0}{c^2}$ \\[4pt]
   $\mathbf{P}_\text{\tiny MP}$       & $\;\;\displaystyle\int\Big(\rho_\mathrm{a}\mathbf{v}_\mathrm{a} + \frac{\mathbf{E}\times\mathbf{H}}{c^2}\Big)d^3r\;\;$ & $\displaystyle\frac{n_\mathrm{p}N_0\hbar\omega_0}{c}\hat{\mathbf{x}}$ \\[4pt]
   $\mathbf{P}_\text{field}$      & $\displaystyle\int\frac{\mathbf{E}\times\mathbf{H}}{c^2}d^3r$ & $\displaystyle\frac{N_0\hbar\omega_0}{n_\mathrm{g}c}\hat{\mathbf{x}}$ \\[4pt]
   $\;\;\mathbf{P}_\text{\tiny MDW}\;\;$             & $\displaystyle\int\rho_\mathrm{a}\mathbf{v}_\mathrm{a}d^3r$ & $\;\;\displaystyle\Big(n_\mathrm{p}-\frac{1}{n_\mathrm{g}}\Big)\frac{N_0\hbar\omega_0}{c}\hat{\mathbf{x}}\;\;$ \\[4pt]
   \hline\hline
 \end{tabular}
\end{table}

\begin{figure}
\centering
\includegraphics[width=0.9\columnwidth]{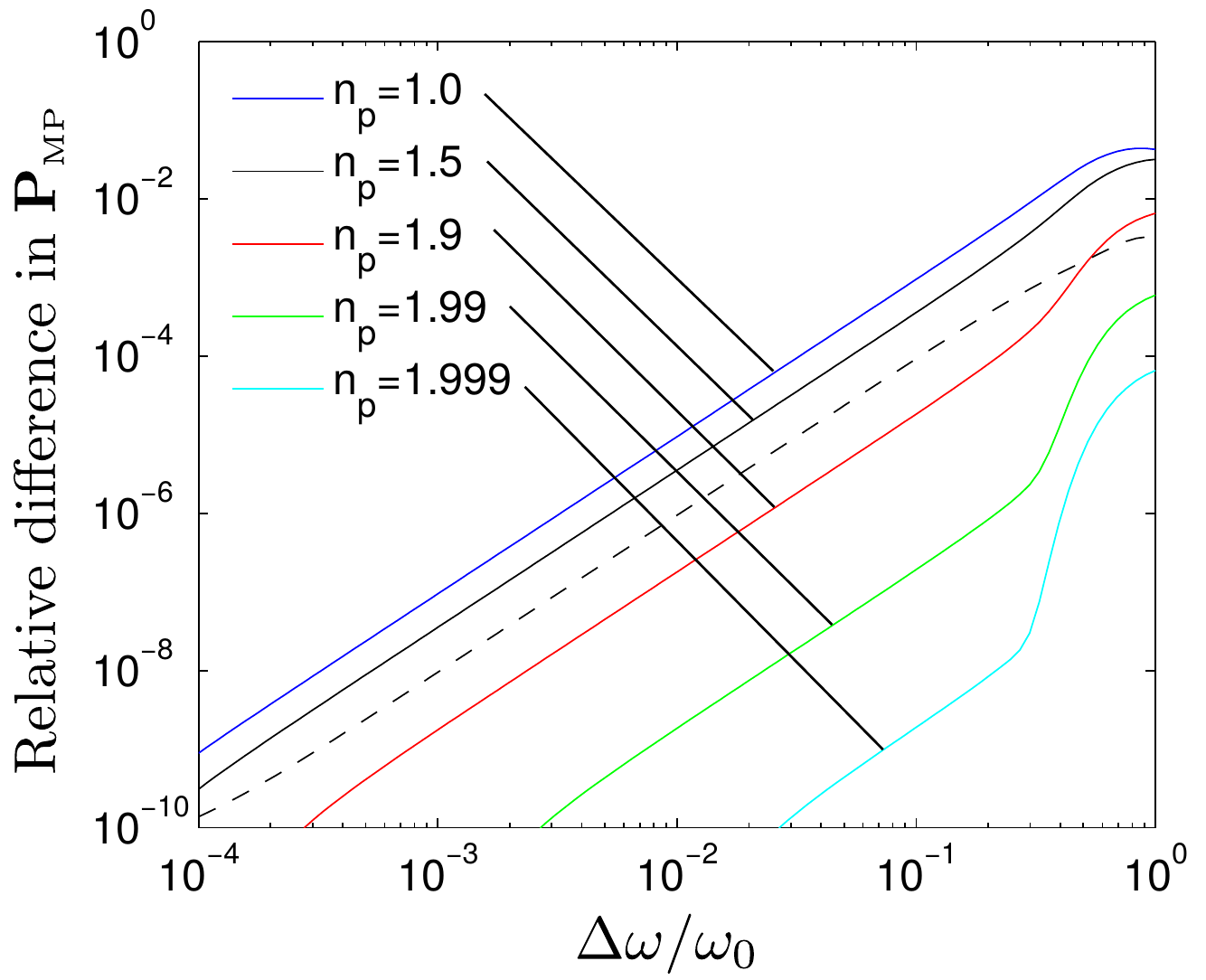}
\vspace{-0.2cm}
\caption{\label{fig:rdp}
(Color online) Relative difference in $\mathbf{P}_\text{\tiny MP}$ obtained from the MP and OCD models
as a function of the relative spectral width of a Gaussian light pulse.
The wavelength is fixed to $\lambda_0=1550$ nm
and the group refractive index is fixed to $n_\mathrm{g}=2$.
Solid lines show the relative difference calculated by using
the linear dispersion relation for selected values of the phase
refractive index. The dashed line shows the corresponding
plot assuming the nonlinearly dispersive material studied in Sec.~\ref{sec:simulationsvisualization},
where $n_\mathrm{p}=1.5$ and $n_\mathrm{g}=2$ for the central frequency.}
\vspace{-0.2cm}
\end{figure}

In order to study the correspondence between the MP and
OCD results in a dispersive medium, we plot the relative difference in the total momentum
$\mathbf{P}_\text{\tiny MP}$ obtained
from the MP and OCD models as a function of the relative spectral
width $\Delta \omega/\omega_0$ of a Gaussian light pulse.
The relative difference is plotted in Fig.~\ref{fig:rdp} for selected
values of the phase refractive index of a linearly dispersive material
when the group refractive index
is fixed to $n_\mathrm{g}=2$ and $\lambda_0=1550$ nm.

Figure \ref{fig:rdp} shows that, for dispersive media, the OCD and MP momenta
are not equal but their relative difference depends critically on
the pulse width. For dispersive media,
the MP and OCD results become equal only
in the narrow band limit $\Delta \omega/\omega_0\rightarrow 0$
for which the relative difference in $\mathbf{P}_\text{\tiny MP}$ becomes zero.
For relative difference values smaller than $10^{-10}$,
the graphs saturate to a constant value (not shown) following from the
accuracy of the numerical simulation.
It is also found that the relative difference in the transferred
mass obtained by using the MP and OCD models behaves the same
way as the relative difference in the total momentum in Fig.~\ref{fig:rdp}.
At the narrow band limit, both models describe very accurately the same
electromagnetic pulse. Thus, one may argue, in analogy to the nondispersive case \cite{Partanen2017c},
that since the MP and OCD models are representations of the same covariant theory for
the single-photon and continuum fields, respectively, the results predicted by them
must be equal.

That the results of the MP and OCD models are not exactly equal for
a broader pulse can be understood by looking at the expressions
of the electric and magnetic fields of the pulse, see Eqs.~\eqref{eq:electricfieldgeneral}
and \eqref{eq:magneticfieldgeneral}, and in particular, keeping in mind
the dispersion relation and the related $k$ dependence of the Fourier
components $\tilde E(k)e^{-i\omega(k)t}$ and $\tilde H(k)e^{-i\omega(k)t}$.
Thus, the OCD model effectively accounts for the frequency dependence of the refractive
index. Accordingly, the OCD result cannot be exactly equal to the MP result if
we multiply the MP total momentum by photon number of the pulse
and keep the refractive indices fixed to their central frequency values.
Note also that the accuracy of the total momentum given by the OCD model is limited
by the fact that the electromagnetic energy density formula
in Eq.~\eqref{eq:fieldenergy} used in the OCD model is known to hold
exactly only in the limit of a monochromatic field \cite{Landau1984}.
Applying the OCD model with a more accurate form of the energy
density given in Ref.~\cite{Philbin2011} is a topic of further work.

The dashed line in Fig.~\ref{fig:rdp} represents the result
for the nonlinearly dispersive case studied in
Sec.~\ref{sec:simulationsvisualization}, where $n_\mathrm{p}=1.5$
and $n_\mathrm{g}=2$ for the central frequency.
One can see that this graph is qualitatively similar to the graphs obtained
for linearly dispersive media. This indicates that the nonlinearity
of the dispersion relation does not significantly influence the
relative difference between the MP and OCD results.

\vspace{-0.2cm}

\subsection{\label{sec:comparisonwithprevious}Comparison of the MP and OCD models with previous experiments and theories}

Neither previous experiments or theories have determined
the transferred mass of the light pulse, but we can still
compare the total MP and OCD momenta of the light pulse
with previous works.
In the narrow-band limit, both the MP and OCD models
give the total momentum which is of the Minkowski
form $p_\text{\tiny MP}=n_\mathrm{p}N_0\hbar\omega_0/c$.
This result is in accordance with the laser beam experiment
of Jones and Leslie \cite{Jones1978}, which as discussed in
Sec.~\ref{sec:experiments} supports the Minkowski formula.
The experiments of Jones and Leslie were carried out for
transparent liquids for which the relaxation dynamics follows
Navier-Stokes equation instead of the elasticity theory.
However, the relaxation dynamics has extremely small influence
on the total momentum of the light pulse.
Thus, our simulations for solids can also explain
the results of Jones and Leslie.
For a reference of possible future experiments making
use of a broad pulse, one should note that the OCD model gives
a more accurate total momentum. The conventional definition
of the Minkowski momentum $p_\text{\tiny MP}=n_\mathrm{p}N_0\hbar\omega_0/c$
is not meaningful since, instead of a constant $n_\mathrm{p}$,
we should use a phase index that is appropriately averaged
over the broad band pulse.

Previously, Garrison \emph{et al.}~\cite{Garrison2004}
have encountered into a problem that the single-photon expectation
value of the momentum $p_\mathrm{M}=\int\mathbf{g}_\mathrm{M}d^3r$,
obtained by using the Minkowski momentum density $\mathbf{g}_\mathrm{M}=\mathbf{D}\times\mathbf{B}$,
is not equal to the commonly defined Minkowski
momentum $\tilde p_\mathrm{M}=n_\mathrm{p}\hbar\omega/c$
following from the de Broglie hypothesis or from the present theory but rather
$p_\mathrm{M}=n_\mathrm{p}^2\hbar\omega/(n_\mathrm{g}c)$.
This controversy has prompted Barnett
to introduce different kinetic and canonical values for the
momentum of light \cite{Barnett2010b,Barnett2010a}. In Barnett's theory, 
$\mathbf{g}_\mathrm{M}$ is called as the canonical momentum density, but the
single-photon value of the momentum is determined by the spatial shift of the field
rather than by the single-photon expectation value of $\int\mathbf{g}_\mathrm{M}d^3r$
\cite{Barnett2010b,Barnett2010a}.
In contrast to Barnett's theory, in our MP theory, all complications are avoided as the
Minkowski momentum $p_\text{\tiny MP}=n_\mathrm{p}\hbar\omega/c$ of the MP is also obtained for the single-photon
expectation value calculated by using the total MP momentum density
$\mathbf{g}_\text{\tiny MP}=\rho_\mathrm{a}\mathbf{v}_\mathrm{a}+\mathbf{E}\times\mathbf{H}/c^2$.

In the literature, there exist also other attempts to explain the difference between
the two different forms of the Minkowski momentum. According to Philbin \cite{Philbin2011}
and the very recent works by Bliokh \emph{et al.}~\cite{Bliokh2017a,Bliokh2017b}, the form
$p_\mathrm{M}=n_\mathrm{p}^2\hbar\omega/(n_\mathrm{g}c)$ is obtained if one neglects certain
dispersion-related terms in the momentum density while accounting for these terms gives the correct
form $\tilde p_\mathrm{M}=n_\mathrm{p}\hbar\omega/c$, which equals the MP momentum
of the present work in Eq.~\eqref{eq:pmp}. These works did not however present
any general splitting of the total Minkowski momentum to the field and the atomic MDW parts.
In the studies of surface plasmon
polaritons (SPPs) \cite{Bliokh2017a,Bliokh2017b}, Bliokh \emph{et al.}
found that there is a current of electrons which accounts for the
difference between the Abraham and Minkowski momenta. Therefore, one can expect that,
in structures supporting SPPs, the MDW may correspond to the excess mass density
due to the moving electrons. The concept of the covariant state of light and the related
MDW are very general and expected to apply to the description of light propagation
in any material structures.

\subsection{Expression of the optical force}

In the derivation of the covariant description of the light pulse,
we have emphasized the coupled state of the field and matter
as an internal property of the MP. 
Also, the OCD simulations are carried far from the interfaces
so that the first term of the optical force density Eq.~\eqref{eq:opticalforcedensity}
including the gradient of the phase refractive
index does not influence the momentum or the transferred mass of the light pulse.
For this internal coupled state, we can forget various
interface effects and consider only the second term
of the optical force density in Eq.~\eqref{eq:opticalforcedensity}.
The second term in Eq.~\eqref{eq:opticalforcedensity} represents
the generalization of the well-known Abraham force for a
dispersive medium.

The full agreement between the MP and OCD models is
obtained only if the Abraham force used in the OCD
model is of the form given in Eq.~\eqref{eq:opticalforcedensity}.
The interface term of the optical force then also obtains an
unambiguous form because of the momentum conservation
at the interfaces. Thus, the MP theory of light in a dispersive
medium leads to a unique expression for the optical force.

\vspace{-0.1cm}

\section{\label{sec:conclusions}Conclusions}

\vspace{-0.1cm}

We have generalized the recently developed MP theory of light for dispersive media
assuming that the absorption and scattering losses are very small.
The total momentum and the transferred mass of the light pulse were derived
both using the MP quasiparticle model and the OCD model.
In the OCD simulations, we have considered only solid dispersive media.
However, the relation of the calculated momentum and the transferred mass to
the phase and group refractive indices and to the pulse shape
is very accurately the same for dispersive liquids.
For liquids, the relaxation dynamics restoring
the equilibrium in the medium is governed by the Navier-Stokes
dynamics instead of the elasticity theory.
The effect of the restoring force on the MP state of light
is evidently very small and it is mainly related to the strain field losses.
Detailed study of the MP theory of light in liquids is a topic of further work.
The MP theory as formulated in this work and in Ref.~\cite{Partanen2017c}
also applies for gases. There, in the OCD analysis, the relaxation
dynamics is described by the Fokker-Planck equation. Note that the
MP quasiparticle model appears to be universally valid
independently on the phase of the medium.
If we know the electromagnetic energy of the field and
the velocity of light in the medium or,
in the dispersive case, the dispersion relation of the medium,
we can independently of the phase of the medium use
the results in Table \ref{tbl:table} to obtain
the total momentum and the transferred mass of the light pulse.

In analogy to the case of a nondispersive medium,
we found an unambiguous correspondence between
the MP quasiparticle and OCD models.
The present results generalize our ultimate solution
of the Abraham-Minkowski controversy
for a dispersive medium.
The interesting feature in the MP theory of light
in a dispersive medium is the atomic MDW
and the related mass transfer which explains
how the total momentum of light is shared between
the field and medium atoms which move under
the influence of the optical force.
The mass transferred by the MDW makes our theory
to fulfill Newton's first law and the covariance condition
of the special theory of relativity.
We have also proven that the covariance condition and
the conservation laws jointly determine the expression
of the optical force on the medium associated with a light pulse.

In analogy to our analysis of light propagation in a nondispersive medium,
we also found in this work that the field and the MDW can
be described using classical variables of the field and medium dynamics.
This implies that the dynamical state of the medium is described
entirely using phase phase, i.e., momentum and position related to each
degree of freedom of the system. Therefore, since momentum and position
are in classical physics unambiguously measurable, in principle to the
desired degree of accuracy, we can experimentally determine at any
moment how the momentum is shared between the field and the medium. This
result is in contrast to many previous works on the momentum of light in a
medium \cite{Pfeifer2007}.

To facilitate the planning of measurements, we
have also carried out simulations of how the displacement
of atoms due to the MDW can be measured in a simple silicon
waveguide structure. In these simulations, we have paid particular
attention in the irradiance breakdown threshold of silicon,
which is one of the main limiting factors in possible experimental setups
as the electromagnetic energy density in the medium cannot be made
arbitrarily large. The OCD model also allows for more
detailed simulations accounting for the waveguide dispersion
and the spreading of the pulse energy in the transverse direction.
These simulations are left as a topic of further work.

\begin{acknowledgments}
This work has in part been funded by the Academy of Finland under contract number 287074 and the Aalto Energy Efficiency Research Programme.
\end{acknowledgments}


\begin{thebibliography}{10}
\newcommand{\enquote}[1]{``#1''}

\bibitem{Partanen2017c}
M.~Partanen, T.~H\"ayrynen, J.~Oksanen, and J.~Tulkki, \enquote{Photon mass
  drag and the momentum of light in a medium,} \emph{Phys. Rev. A} \textbf{95},
  063850 (2017).

\bibitem{Landau1984}
L.~D. Landau, E.~M. Lifshitz, and L.~P. Pitaevskii, \emph{Electrodynamics of
  Continuous Media}, Pergamon, Oxford (1984).

\bibitem{Leonhardt2006}
U.~Leonhardt, \enquote{Momentum in an uncertain light,} \emph{Nature}
  \textbf{444}, 823 (2006).

\bibitem{Cho2010}
A.~Cho, \enquote{Century-long debate over momentum of light resolved?}
  \emph{Science} \textbf{327}, 1067 (2010).

\bibitem{Pfeifer2007}
R.~N.~C. Pfeifer, T.~A. Nieminen, N.~R. Heckenberg, and H.~Rubinsztein-Dunlop,
  \enquote{Colloquium: Momentum of an electromagnetic wave in dielectric
  media,} \emph{Rev. Mod. Phys.} \textbf{79}, 1197 (2007).

\bibitem{Barnett2010b}
S.~M. Barnett, \enquote{Resolution of the {A}braham-{M}inkowski dilemma,}
  \emph{Phys. Rev. Lett.} \textbf{104}, 070401 (2010).

\bibitem{Barnett2010a}
S.~M. Barnett and R.~Loudon, \enquote{The enigma of optical momentum in a
  medium,} \emph{Phil. Trans. R. Soc. A} \textbf{368}, 927 (2010).

\bibitem{Leonhardt2014}
U.~Leonhardt, \enquote{Abraham and {M}inkowski momenta in the optically induced
  motion of fluids,} \emph{Phys. Rev. A} \textbf{90}, 033801 (2014).

\bibitem{Saldanha2017}
P.~L. Saldanha and J.~S.~O. Filho, \enquote{Hidden momentum and the
  {A}braham-{M}inkowski debate,} \emph{Phys. Rev. A} \textbf{95}, 043804
  (2017).

\bibitem{Brevik2017}
I.~Brevik, \enquote{Minkowski momentum resulting from a vacuum-medium mapping
  procedure, and a brief review of {M}inkowski momentum experiments,}
  \emph{Ann. Phys.} \textbf{377}, 10  (2017).

\bibitem{Ramos2015}
T.~Ramos, G.~F. Rubilar, and Y.~N. Obukhov, \enquote{First principles approach
  to the {A}braham-{M}inkowski controversy for the momentum of light in general
  linear non-dispersive media,} \emph{J. Opt.} \textbf{17}, 025611 (2015).

\bibitem{Crenshaw2013}
M.~E. Crenshaw, \enquote{Decomposition of the total momentum in a linear
  dielectric into field and matter components,} \emph{Ann. Phys.}
  \textbf{338}, 97 (2013).

\bibitem{Mansuripur2010}
M.~Mansuripur, \enquote{Resolution of the {A}braham-{M}inkowski controversy,}
  \emph{Opt. Commun.} \textbf{283}, 1997  (2010).

\bibitem{Kemp2011}
B.~A. Kemp, \enquote{Resolution of the {A}braham-{M}inkowski debate:
  Implications for the electromagnetic wave theory of light in matter,}
  \emph{J. Appl. Phys.} \textbf{109}, 111101 (2011).

\bibitem{Abraham1909}
M.~Abraham, \enquote{Zur {E}lektrodynamik bewegter {K}\"orper,} \emph{Rend.
  Circ. Matem. Palermo} \textbf{28}, 1 (1909).

\bibitem{Abraham1910}
M.~Abraham, \enquote{Sull'elettrodinamica di {M}inkowski,} \emph{Rend. Circ.
  Matem. Palermo} \textbf{30}, 33 (1910).

\bibitem{Minkowski1908}
H.~Minkowski, \enquote{Die {G}rundgleichungen f\"ur die elektromagnetischen
  {V}org\"ange in bewegten {K}\"orpern,} \emph{Nachr. Ges. Wiss. G\"ottn
  Math.-Phys. Kl.} 53 (1908), reprinted in Math. Ann. \textbf{68}, 472 (1910).

\bibitem{Campbell2005}
G.~K. Campbell, A.~E. Leanhardt, J.~Mun, M.~Boyd, E.~W. Streed, W.~Ketterle,
  and D.~E. Pritchard, \enquote{Photon recoil momentum in dispersive media,}
  \emph{Phys. Rev. Lett.} \textbf{94}, 170403 (2005).

\bibitem{Sapiro2009}
R.~E. Sapiro, R.~Zhang, and G.~Raithel, \enquote{Atom interferometry using
  {K}apitza-{D}irac scattering in a magnetic trap,} \emph{Phys. Rev. A}
  \textbf{79}, 043630 (2009).

\bibitem{Jones1954}
R.~V. Jones and J.~C.~S. Richards, \enquote{The pressure of radiation in a
  refracting medium,} \emph{Proc. R. Soc. Lond. A} \textbf{221}, 480 (1954).

\bibitem{Jones1978}
R.~V. Jones and B.~Leslie, \enquote{The measurement of optical radiation
  pressure in dispersive media,} \emph{Proc. R. Soc. Lond. A} \textbf{360}, 347
  (1978).

\bibitem{Walker1975}
G.~B. Walker and D.~G. Lahoz, \enquote{Experimental observation of {A}braham
  force in a dielectric,} \emph{Nature} \textbf{253}, 339 (1975).

\bibitem{She2008}
W.~She, J.~Yu, and R.~Feng, \enquote{Observation of a push force on the end
  face of a nanometer silica filament exerted by outgoing light,} \emph{Phys.
  Rev. Lett.} \textbf{101}, 243601 (2008).

\bibitem{Zhang2015}
L.~Zhang, W.~She, N.~Peng, and U.~Leonhardt, \enquote{Experimental evidence for
  {A}braham pressure of light,} \emph{New J. Phys.} \textbf{17}, 053035 (2015).

\bibitem{Astrath2014}
N.~G.~C. Astrath, L.~C. Malacarne, M.~L. Baesso, G.~V.~B. Lukasievicz, and
  S.~E. Bialkowski, \enquote{Unravelling the effects of radiation forces in
  water,} \emph{Nat. Commun.} \textbf{5}, 4363 (2014).

\bibitem{Ashkin1973}
A.~Ashkin and J.~M. Dziedzic, \enquote{Radiation pressure on a free liquid
  surface,} \emph{Phys. Rev. Lett.} \textbf{30}, 139 (1973).

\bibitem{Casner2001}
A.~Casner and J.-P. Delville, \enquote{Giant deformations of a liquid-liquid
  interface induced by the optical radiation pressure,} \emph{Phys. Rev. Lett.}
  \textbf{87}, 054503 (2001).

\bibitem{Brevik1979}
I.~Brevik, \enquote{Experiments in phenomenological electrodynamics and the
  electromagnetic energy-momentum tensor,} \emph{Phys. Rep.} \textbf{52}, 133
  (1979).

\bibitem{Griffiths1998}
D.~J. Griffiths, \emph{Introduction to Electrodynamics}, Prentice-Hall, Upper
  Saddle River, NJ (1998).

\bibitem{Milonni2010}
P.~W. Milonni and R.~W. Boyd, \enquote{Momentum of light in a dielectric
  medium,} \emph{Adv. Opt. Photon.} \textbf{2}, 519 (2010).

\bibitem{Peiponen1999}
K.-E. Peiponen, E.~M. Vartiainen, and T.~Asakura, \emph{Dispersion, Complex
  Analysis and Optical Spectroscopy: Classical Theory}, Springer, Berlin
  (1999).

\bibitem{Bedford1994}
A.~Bedford and D.~S. Drumheller, \emph{Introduction to Elastic Wave
  Propagation}, Wiley, Chichester (1994).

\bibitem{Mavko2003}
G.~Mavko, T.~Mukerji, and J.~Dvorkin, \emph{The Rock Physics Handbook},
  Cambridge University Press, Cambridge (2003).

\bibitem{Kittel2005}
C.~Kittel, \emph{Introduction to Solid State Physics}, Wiley, Hoboken, NJ
  (2005).

\bibitem{Philbin2011}
T.~G. Philbin, \enquote{Electromagnetic energy momentum in dispersive media,}
  \emph{Phys. Rev. A} \textbf{83}, 013823 (2011).

\bibitem{Garrison2004}
J.~C. Garrison and R.~Y. Chiao, \enquote{Canonical and kinetic forms of the
  electromagnetic momentum in an \textit{ad hoc} quantization scheme for a
  dispersive dielectric,} \emph{Phys. Rev. A} \textbf{70}, 053826 (2004).

\bibitem{ChenJi2011}
J.~Chen, Y.~Wang, B.~Jia, T.~Geng, X.~Li, L.~Feng, W.~Qian, B.~Liang, X.~Zhang,
  M.~Gu, and S.~Zhuang, \enquote{Observation of the inverse {D}oppler effect in
  negative-index materials at optical frequencies,} \emph{Nat. Photon.}
  \textbf{5}, 239 (2011).

\bibitem{Milonni2005}
P.~W. Milonni and R.~W. Boyd, \enquote{Recoil and photon momentum in a
  dielectric,} \emph{Laser Phys.} \textbf{15}, 1432 (2005).

\bibitem{Smith2007}
A.~Smith, B.~Do, and M.~Soderlund, \enquote{Deterministic nanosecond
  laser-induced breakdown thresholds in pure and {Y}b$^{3+}$ doped fused
  silica,} \emph{Proc. SPIE} \textbf{6453}, 645317 (2007).

\bibitem{supplementaryvideo}
See Supplemental Material at http://link.aps.org/\linebreak
supplemental/10.1103/PhysRevA.96.063834 for video files
  of the simulated {MDW} and atomic displacements in linearly and nonlinearly
  dispersive media.

\bibitem{Li1980}
H.~H. Li, \enquote{Refractive index of silicon and germanium and its wavelength
  and temperature derivatives,} \emph{J. Phys. Chem. Ref. Data} \textbf{9}, 561
  (1980).

\bibitem{Lide2004}
D.~R. Lide, ed., \emph{CRC Handbook of Chemistry and Physics}, CRC Press, Boca
  Raton, FL (2004).

\bibitem{Hopcroft2010}
M.~A. Hopcroft, W.~D. Nix, and T.~W. Kenny, \enquote{What is the {Y}oung's
  modulus of silicon?} \emph{J. Microelectromech. Syst.} \textbf{19}, 229
  (2010).

\bibitem{Cowan2006}
B.~Cowan, \enquote{Optical damage threshold of silicon for ultrafast infrared
  pulses,} \emph{AIP Conf. Proc.} \textbf{877}, 837 (2006).

\bibitem{Schinke2015}
C.~Schinke, P.~C. Peest, J.~Schmidt, R.~Brendel, K.~Bothe, M.~R. Vogt,
  I.~Kr\"oger, S.~Winter, A.~Schirmacher, S.~Lim, H.~T. Nguyen, and
  D.~MacDonald, \enquote{Uncertainty analysis for the coefficient of
  band-to-band absorption of crystalline silicon,} \emph{AIP Adv.} \textbf{5},
  067168 (2015).

\bibitem{Green2008}
M.~A. Green, \enquote{Self-consistent optical parameters of intrinsic silicon
  at 300{K} including temperature coefficients,} \emph{Sol. Energ. Mat. Sol.
  Cells} \textbf{92}, 1305  (2008).

\bibitem{Bliokh2017a}
K.~Y. Bliokh, A.~Y. Bekshaev, and F.~Nori, \enquote{Optical momentum, spin, and
  angular momentum in dispersive media,} \emph{Phys. Rev. Lett.} \textbf{119},
  073901 (2017).

\bibitem{Bliokh2017b}
K.~Y. Bliokh, A.~Y. Bekshaev, and F.~Nori, \enquote{Optical momentum and
  angular momentum in complex media: from the {A}braham-{M}inkowski debate to
  unusual properties of surface plasmon-polaritons,} \emph{New J. Phys.}
  \textbf{19}, 123014 (2017).

\end{thebibliography}
\end{document}